\documentclass[12pt]{article}
\usepackage{latexsym}
\usepackage{amsfonts}
\usepackage{epsfig}

\begin{document}
\def\op#1{\mathcal{#1}}
\def\eq#1{(\ref{#1})}

\def\IC{\relax\,\hbox{$\inbar\kern-.3em{\rm C}$}}
\def\inbar{\vrule height1.5ex width.4pt depth0pt}

\def\a{\alpha}
\def\b{\beta}
\def\bfone{\relax{\rm 1\kern-.35em 1}}
\def\bfnull{\relax{\rm O \kern-.635em 0}}
\def\bos{{\rm bos}}
\def\c{\chi}
\def\cb{\bar{\chi}}
\def\cc{{\rm c.c.}}
\def\d{\delta}
\def\D{\Delta}
\def\der{\partial}
\def\dop{{\rm d}\hskip -1pt}
\def\dx{\right}
\def\e{\epsilon}
\def\g{\gamma}
\def\G{\Gamma}
\def\i{\imath}
\def\im{{\rm Im}\op{N}}
\def\imez{\frac{{\rm i}}{2}}
\def\l{\lambda}
\def\L{\Lambda}
\def\m{\mu}
\def\mez{\frac{1}{2}}
\def\n{\nu}
\def\na{\nabla}
\def\o{\omega}
\def\O{\Omega}
\def\ol{\overline}
\def\p{\psi}
\def\qu{\frac{1}{4}}
\def\r{\rho}
\def\re{{\rm Re}\op{N}}
\def\s{\sigma}
\def\S{\Sigma}
\def\sx{\left}
\def\t{\tau}
\def\th{\theta}
\def\Th{\Theta}
\def\ve{\varepsilon}
\def\z{\zeta}

\newcommand{\be}{\begin{equation}}
\newcommand{\ee}{\end{equation}}
\newcommand{\ba}{\begin{eqnarray}}
\newcommand{\ea}{\end{eqnarray}}
\newcommand{\ban}{\begin{eqnarray*}}
\newcommand{\ean}{\end{eqnarray*}}
\newcommand{\nn}{\nonumber}
\newcommand{\noi}{\noindent}
\newcommand{\fgl}{\mathfrak{gl}}
\newcommand{\fu}{\mathfrak{u}}
\newcommand{\fsl}{\mathfrak{sl}}
\newcommand{\fsp}{\mathfrak{sp}}
\newcommand{\fusp}{\mathfrak{usp}}
\newcommand{\fsu}{\mathfrak{su}}
\newcommand{\fp}{\mathfrak{p}}
\newcommand{\fso}{\mathfrak{so}}
\newcommand{\fg}{\mathfrak{g}}
\newcommand{\fr}{\mathfrak{r}}
\newcommand{\fe}{\mathfrak{e}}
\newcommand{\rE}{\mathrm{E}}
\newcommand{\rSp}{\mathrm{Sp}}
\newcommand{\rSO}{\mathrm{SO}}
\newcommand{\rSL}{\mathrm{SL}}
\newcommand{\rSU}{\mathrm{SU}}
\newcommand{\rUSp}{\mathrm{USp}}
\newcommand{\rU}{\mathrm{U}}
\newcommand{\rF}{\mathrm{F}}
\newcommand{\R}{\mathbb{R}}
\newcommand{\C}{\mathbb{C}}
\newcommand{\Z}{\mathbb{Z}}
\newcommand{\Hb}{\mathbb{H}}
\def\N\mathcal{N}

%%%%%%%%%%%%%%%%%%%%%%%%%%%%%%%%%%%%%%%%%%%%%%%%%%%%%%%%%%%%%%%%%%%%%%%%
%%%%%%%%%%%%%%%%%%%%%%%%%%%%%%%%%%%%%%%%%%%%%%%%%%%%%%%%%%%%%%%%%%%%
%%%%%%%%%%%%%%%%%%%%%% FINE MANIFOLDWALKER MACRO %%%%%%%%%%%%%%%%%%%%
%%%%%%%%%%%%%%%%%%%%%%%%%%%%%%%%%%%%%%%%%%%%%%%%%%%%%%%%%%%%%%%%%%%%%%%

\begin{titlepage}

\begin{center}
{\LARGE { $N=2$ Supergravity Lagrangian Coupled to Tensor Multiplets with  Electric and Magnetic Fluxes}}\\
\vskip 1.5cm
  {\bf Riccardo D'Auria$^{a,1}$, Luca Sommovigo$^{a,2}$ and
  Silvia Vaul\`a$^{b,3}$} \\
\vskip 0.5cm
\end{center}
\begin{center}
{\small $^{a}$ Dipartimento di Fisica, Politecnico di
Torino,\\ Corso Duca degli Abruzzi 24, I-10129 Torino, Italy}\\
{\small and}\\
{\small  Istituto Nazionale di Fisica Nucleare (INFN) - Sezione di
Torino,\\ Via P. Giuria 1, I-10125 Torino, Italy\\
\vskip 0.5cm
$^{b}$ DESY, Theory Group\\
Notkestrasse 85, Lab. 2a D-22603 Hamburg, Germany}
\end{center}
\begin{center}
e-mail: $^{1}$riccardo.dauria@polito.it\quad
$^{2}$luca.sommovigo@polito.it\\  $^{3}$silvia.vaula@desy.de
\end{center}
\vskip 1cm
\begin{abstract}

We derive the full $N=2$ supergravity Lagrangian which contains a
symplectic invariant scalar potential in terms of electric and
magnetic charges. As shown in reference \cite{Dall'Agata:2003yr},
the appearance of magnetic charges is allowed only if tensor
multiplets are present and a suitable Fayet--Iliopoulos term is
included in the fermion transformation laws. We generalize the
procedure in the quoted reference by adding further a
Fayet--Iliopoulos term which allows the introduction of electric
charges in such a way that the potential and the equations of
motion of the theory are symplectic invariant. The theory is
further generalized to include an ordinary electric gauging and
the form of the resulting scalar potential is given.

\end{abstract}

\end{titlepage}

\section{Introduction}
\label{sec:intro}

Standard $D=4$ supergravities with $N$ supersymmetries are usually
described, for $N\leq 4$, as the coupling of the gravitational
multiplet to matter multiplets whose bosonic content is limited to
scalar or vector fields. Multiplets containing massless
twofold--antisymmetric tensor fields are usually converted into
ordinary vector or scalar multiplets using Hodge duality in four
dimensions between tensors and scalars \cite{deWit:1982na}.
However the description of the couplings and the geometry of the
$\sigma$--model is quite different in the dual theory. This is the
case, for example, of the coupling of an arbitrary number of
tensor multiplets (linear multiplets) in $N=1$ Supergravity
\cite{Bertolini:1994cb} and of the coupling of the scalar--tensor
multiplet in $N=2$ supergravity \cite{Theis:2003jj}. From the
point of view of compactification of ten dimensional
supergravities down to four dimensions, tensor multiplets appear
naturally in $D=4$ due to the presence of antisymmetric tensors in
the $D=10$ theory. For a full understanding of four dimensional
gauged or ungauged supergravities in the presence of tensor field
fluxes in the parent ten dimensional theories, it seems
appropriate to keep at least some of the massive tensor fields
undualized in the
resulting four dimensional Lagrangian.\\
In reference \cite{Louis:2002ny} it has been shown, at a purely
bosonic level, that ten dimensional Type IIA and Type IIB
supergravities compactified on a Calabi--Yau threefold, give rise
to theories with with tensor multiplets and that in the presence
of general fluxes they contain  both electric and magnetic
charges. Particularly, some of the tensors might become massive
due to the presence of the magnetic fluxes. For instance in the
Type IIB case the universal hypermultiplet appears as a double
tensor multiplet $(B^{(1)}_{\mu\nu},B^{(2)}_{\mu\nu},\ell,\phi)$
\cite{Berkovits:1995cb} and one of the two tensors, which come
either from the R--R and NS--NS 3--form respectively, is massive
when the magnetic R--R or NS--NS flux is switched on, due to
following coupling with the vector field strengths:

\begin{equation}
\label{redeff} \op{F}^\L \to {\hat \op{F}}^\L = \op{F}^\L +
2m^{I\L} B_I
\end{equation}

\noi where $I=1$ or $I=2$ according to the case. Furthermore the
theory with both electric and magnetic fluxes has a symplectic
invariant scalar potential and the set of equations of motion and
Bianchi identities for the vector field strengths are also
symplectic invariant as in the ungauged supergravity, provided
that the electric and magnetic fluxes $(m^{I\L},\,e^I_\L)$
transform as a symplectic vector. In reference
\cite{Taylor:1999ii}, the scalar
potential in the presence of both R--R and NS--NS fluxes was derived.\\
This result naturally rises the question of how to understand the
magnetic charges from a purely four dimensional point of view,
without introducing ill--defined magnetic gauge potentials in the
theory, as it was done in reference \cite{Michelson:1996pn}. The
answer was found in reference \cite{Dall'Agata:2003yr} where it
was shown that in $N=2$ supergravity coupled to vector multiplets
and to the $N=2$ scalar--tensor multiplet (tensor multiplet in the
following), the appearance of magnetic charges can be understood
as the consequence of the redefinition of the vector
field-strengths as in equation (\ref{redeff}), where now, however,
the index $I$ runs on the number of antisymmetric tensor fields
contained in the tensor multiplet. This is consistent with
supersymmetry provided one performs a non--trivial generalization
of the fermionic transformation laws of the ungauged theory. This
modification amounts to adding to the transformation laws of the
fermions a magnetic "mass--shift" (or Fayet--Iliopoulos term)
which is therefore allowed if and only if antisymmetric tensor
fields are present.\\
In the quoted reference the theory was also generalized performing
an ordinary electric gauging. However the electric charges
introduced by the gauging cannot be identified with the electric
charges of references \cite{Louis:2002ny},\cite{Taylor:1999ii}
since they are not naturally paired with the magnetic charges
giving rise to a symplectic vector. Indeed, if one thinks of $N=2$
supergravity with tensor multiplets as coming from the dualization
of the axionic fields on the quaternionic manifold $\mathcal{M}_Q$
then one is left with a new manifold $\mathcal{M}_T$ without the
translational symmetries which have been dualized. Therefore one
can not obtain, on the manifold $\mathcal{M}_T$, electric Killing
vectors which are charged under the original axionic symmetries.\\
In reference \cite{Sommovigo:2004vj}, the dualization of the
axionic coordinates is performed after the translational gauging
has been implemented. This way one obtains that the electric
charges $e^I_\L$ associated to the gauging are naturally paired
with the magnetic charges which one can generate by means of a
suitable symplectic rotation. The approach of
\cite{Sommovigo:2004vj} however is not the most efficient in order
to implement supersymmetry and to derive the supersymmetric
Lagrangian, which in fact was not constructed.\\
In this paper we take the attitude of constructing the full
supersymmetric theory, namely the Lagrangian and its supersymmetry
transformation laws, directly from the content of the
gravitational, vector, and tensor supermultiplets, thus using as a
$\s$--model the manifold $\mathcal{M}_T \otimes \mathcal{M}_{SK}$.
Here $\op{M}_T$ is the $\s$--model parametrized by the scalar
fields of the $N=2$ tensor multiplet, and $\mathcal{M}_{SK}$ is
the Special K\"ahler manifold of the vector multiplets (which
remains untouched with respect to the standard $N=2$
supergravity). Our results show that it is possible to have
electric charges naturally paired with the magnetic ones without
performing any translational gauging, but simply adding, to the
transformation laws of the fermions, a further Fayet--Iliopoulos
term or "electric mass--shift", besides the magnetic one of
reference \cite{Dall'Agata:2003yr}. As a result of our analysis of
the Bianchi identities in superspace, supersymmetry then implies
that the mass--shifts are symplectic invariant if one assumes that
the vector $\sx( m^{I\L},e^I_\L \dx)$ is symplectic, that is
transforms as the symplectic section of the Special Geometry.
Since the scalar potential is constructed as a quadratic form in
the mass--shifts it is automatically symplectic invariant. As a
particular case the scalar potential obtained in this way
coincides with the scalar potential obtained in the case of
Calabi--Yau compactification from Type IIA and Type IIB
supergravity \cite{Taylor:1999ii}, when the scalar--tensor
multiplets are suitably specified. Therefore, as a particular case
of our construction we obtain the complete four dimensional
supergravity corresponding to Type IIB compactification on CY in
the presence of both R--R and NS--NS electric
and magnetic fluxes.\\
Note that as in the standard ungauged $N=2$ theory the equations
of motion and the Bianchi identities for the vectors are still
covariant under symplectic transformations of the electric and
magnetic field--strengths. If we perform an ordinary gauging of
the theory, then of course the on--shell symplectic invariance is
broken leaving however the symplectic invariance in
the sector parametrized by $m^{I\L}, e^I_\L$.\\
We also note that our Lagrangian is invariant under the combined
gauge transformations \cite{Louis:2002ny}:

\begin{eqnarray}
\d B_{I\m\n} &=& \der_{[\m} \L_{I\n]} \\
\d A_\m^\L &=& - 2 m^{I\L} \L_{I\m}
\end{eqnarray}

\noi where $I$ is the number of 2--forms of the tensor multiplet,
and $\L_{I\m}$ is a 1--form gauge parameter, if and only if the
electric and magnetic charges satisfy the constraint:

\begin{equation}
e^I_\L m^{J\L} - e^J_\L m^{I\L} = 0.
\end{equation}

\noi The same constraint also appears from the supersymmetry Ward
identity for the scalar potential \cite{Sommovigo:2004vj}. This
constraint is a generalization of the tadpole cancellation
mechanism in ten dimensions, where the indices $I,J$ take now only
the
values $1,2$ associated to the NS-NS and R-R 2--form.\\
~\\
\noi The paper is organized as follows:\\
In Section \ref{stsugra} we describe how to generalize the
Fayet--Iliopoulos mechanism related to magnetic charges to include
also a further mass--shift related to the electric charges,
therefore obtaining symplectic invariant expression for the
mass--shifts of the fermions.\\
In Section \ref{Lagrangian} we give the Lagrangian and the
supersymmetry transformation laws for the $N=2$ supergravity
coupled to an arbitrary number of vector multiplets and to the
tensor multiplet in our general setting, where electric and
magnetic charges are present (the method we use is
described in Appendix B).\\
In Subsection \ref{adding} we further generalize the Lagrangian by
gauging  the electric vector potentials with an arbitrary group
$\op{G}$, while in Subsection \ref{discpot} we discuss the general
form of the scalar potential in this more
general setting.\\
In Appendix A we give the completion of the Lagrangian and
supersymmetry transformation laws including 4--fermion and
3--fermion terms respectively which are not present in the main
text.\\
In Appendix B we present the superspace approach for the solution
of the Bianchi identities and the rheonomic approach for the
construction of the Lagrangian.

\section{The $D=4$ $N=2$ Supergravity theory\\ coupled to vector multiplets and
scalar--tensor multiplets} \label{stsugra}
\setcounter{equation}{0}

In reference \cite{Dall'Agata:2003yr}, \cite{Theis:2003jj} the
general ungauged $N=2$ theory in the presence of vector multiplets
and scalar--tensor multiplets was constructed. We recall the basic
definitions. In this theory we have besides the gravitational
multiplet, $n_V$ vector multiplets and $n_H$ hypermultiplets. The
gravitational multiplet contains the graviton $V^a$, the
(anti)--chiral gravitinos ($\psi^A$), $\psi_A$ and the graviphoton
$A^0_\m$, where $A=1,2\,$ is the ${\rm SU}(2)$ $R$--symmetry
index. The $n_V$ vector multiplets contain $n_V$ vector, $2n_V$
(anti)--chiral gauginos ($\l^{\bar \imath}_A$), $\l^{iA}$ and the
complex scalar fields $z^i$, $i=1,\dots ,n_V$ which parametrize a
$2n_V$ dimensional special K\"ahler manifold $\mathcal{M}_{SK}$.
We denote by $A^\L_\m$, $\L= 0, \dots ,n_V$, the graviphoton
($\L=0$) and the $n_V$ vectors of the vector multiplets, where
$\L$ is the symplectic index in the half upper part of the
holomorphic section of the symplectic bundle which defines the
special K\"ahler manifold $\op{M}_{SK}$. To this sector we add a
scalar--tensor multiplet (in the following tensor multiplet) whose
content is: ($B_{I\,\m\n}$, $\z_\a$, $\z^\a$, $q^u$) $I=1,\dots
,n_T$, $u=1 ,\dots, 4n_H-n_T$, $\a=1,\dots ,2n_H$ where ($\z^\a$)
$\z_\a$ are the (anti)--chiral component of the spin $1/2$
fermions. This tensor multiplet can be understood as resulting
from the hypermultiplet sector of standard $N=2$
supergravity\footnote{Here and in the following by standard $N=2$
supergravity we mean $N=2$ supergravity coupled to $n_V$ vector
multiplets and $n_H$ hypermultiplets as formulated in references
\cite{Andrianopoli:1996cm},\cite{Andrianopoli:1996ve}.} by
dualization of the axionic coordinates of the quaternionic
manifold. Indeed, the $n_H$ hypermultiplets, in the standard
formulation contain $2n_H$ (anti)--chiral hyperinos $(\z^\a)$,
$\z_\a$, where $\a=1,\dots ,2n_H$ is a Sp$(2n_H,\mathbb R)$ index,
and $q^{\hat u}$ scalars, $\hat{u}=1,\dots, 4n_H$, which
parametrize a $4n_H$--dimensional quaternionic manifold
$\mathcal{M}_Q$. If the quaternionic manifold $\mathcal{M}_Q$
admits $n_T$ translational isometries, we can split the scalars
$q^{\hat u}=(q^u,q^I)$, $u=1, \dots ,4n_H-n_T$, $I=1,\dots ,n_T$,
where the $q^I$'s are the axionic scalars associated to the
translations that can therefore be dualized into $n_T$ tensors
$B_{I\m\n}$.\\
The $\s$--model of the resulting theory is parametrized by the
coordinates $q^u$ and in the following we will refer to it as
$\op{M}_T$. While in the standard $N=2$ supergravity the
quaternionic manifold has holonomy contained in
SU$(2)\otimes$Sp$(2n_H, \mathbb R)$, $\op{M}_T$ has a reduced
holonomy which is however contained in SU$(2)\otimes$Sp$(2n_H,
\mathbb R)\otimes$SO$(n_T)$.\\
In reference \cite{Dall'Agata:2003yr} it was found that $N=2$
supergravity coupled to $n_V$ vector multiplets and a tensor
multiplet admits a non--trivial extension given by the addition of
Fayet--Iliopoulos terms (or magnetic mass--shifts) which generate
a non--trivial scalar potential. Indeed, in order to satisfy the
requirements of supersymmetry, one finds, by suitable analysis of
superspace Bianchi identities, that the transformation laws of the
fermions may be extended with mass--shifts (Fayet--Iliopoulos
terms) defined by:

\begin{eqnarray}
\d\p_{A\m}&=&\d_0 \p_{A\m} + {\rm i}S_{AB}^{(m)}\g_\m\e^B\label{eq:sab}\\
\d\l^{iA}&=&\d_0 \l^{iA} + W^{iAB (m)}\e_B\label{eq:wab}\\
\d\z_\a&=&\d_0 \z_\a + N_\a^{A(m)}\e_A \label{eq:na}
\end{eqnarray}

\noi where the explicit form of the $\d_0$--shifts can be read
from equations (\ref{trasfgrav0},\break
\ref{gaugintrasfm0},\ref{iperintrasf0}) of section
\ref{Lagrangian} up to 3--fermion terms (the 3--fermion terms are
given in Appendix A) and the "magnetic" mass--shifts are given by:

\begin{eqnarray}
S_{AB}^{(m)}&=&- \imez\sigma^x_{AB}\o^x_I m^{I\L}M_\Lambda
\label{shiftSm}\\
W^{iAB(m)} &=& -{\rm i}g^{i\bar \jmath} \sigma^{x\,AB} \o^x_I
m^{I\L} \ol{h}_{\bar\jmath\Lambda}\label{shiftWm}\\
N^{A(m)}_\a&=&-2\mathcal{U}^A_{I\,\a}m^{I\L}M_\L\label{shiftNm}
\end{eqnarray}

\noi where $m^{I\L}$, $I=1, \dots ,n_T$ and $\L=0, \dots ,n_V$,
are constants with the dimension of a mass and as we will see in
the following, they can be interpreted as magnetic charges.
Furthermore $\o^x_I(q^u)$ and $\op{U}_I{}^{A\a}(q^u)$ are
SO$(n_T)$ vector--valued fields on $\op{M}_T$ carrying additional
indices in the adjoint of SU$(2)$ and in fundamental of ${\rm SU}
(2) \otimes {\rm Sp} (2n,\mathbb R)$, respectively. Finally
$M(z,\ol z)_\L$ and $h(z, \ol z)_\L^j$ are the lower component of
the symplectic sections

\begin{eqnarray}\label{sympsect}
V &=& (L^\L,M_\L)\\
U_i &=& D_i V = (f^\L_i, h_{\L i})
\end{eqnarray}

\noi of the special geometry, which transform covariantly under
the $(2 n_V+2) \times (2n_V+2)$ symplectic matrix:

\begin{equation}\label{sympmat}
S = \begin{pmatrix}{A&B\cr C&D}\end{pmatrix}
\end{equation}

\noi where

\begin{equation}
A^T C - C^T A = B^T D - D^T B =0, \quad A^T D - C^T B = \bfone
\end{equation}

\noi It is important to note that, in order to have a
supersymmetric theory, the magnetic mass--shift deformations imply
that the vector field--strengths have to be redefined as follows
\cite{Louis:2002ny},\cite{Romans:1986tw},\cite{D'Auria:2000ad}:

\begin{equation}
\label{redeff1} \op{F}^\L \to {\hat \op{F}}^\L = \op{F}^\L +
2m^{I\L} B_I.
\end{equation}

\noi From equation (\ref{redeff}) it follows that the resulting
theory must contain massive tensor fields $B_{I\m\n}$ since by a
Higgs mechanism a subset (for $n_T<n_V +1$) of the vector
potentials are eaten giving mass to the tensors.\\
Given the mass--shifts, the scalar potential can be computed from
the usual Ward identity of supersymmetry and turns out to be:

\begin{eqnarray}
\label{potential} \op{V} &=& -\mez \o^x_I \o^x_J \sx( {\rm
Im}\op{N}_{\L\S}+ \sx( {\rm Re} \, \op{N} {\rm Im}\op{N}^{-1}\,
{\rm
Re}\op{N}\dx)_{\L\S} \dx) m^{I\L} m^{J\S} +\nn\\
&&+ 4 m^{I\L} m^{J\S} \ol M_\L M_\S \sx(\op{M}_{IJ} - \o^x_I
\o^x_J \dx)
\end{eqnarray}

\noi where $\op{N}_{\L\S}$ is the period matrix of the special
geometry, and the dependence on the coordinates of $\op{M}_T$
appears only in $\o^x_I(q)$.\\
If one thinks of this supergravity theory as coming from standard
$N=2$ supergravity where dualization of the axionic coordinates of
the quaternionic manifold has been performed, one can easily see
that $\o^x_I$ and $\op{U}_I{}^{A\a}$ are the remnants of the ${\rm
SU}(2)$ connection and vielbeins of the quaternionic manifold in
the directions $I$ of the dualized coordinates $q^I$
\cite{Dall'Agata:2003yr}. Moreover the mass parameters $m^{I\L}$
can be associated to the dual algebra of the translation group
isometries on the quaternionic manifold and could be interpreted
as magnetic charges allowed by the presence of the tensor
multiplets \cite{Dall'Agata:2003yr}, \cite{Sommovigo:2004vj}.\\
In fact the potential (\ref{potential}), in the case of
Calabi--Yau compactification of type IIB string theory
\cite{Sommovigo:2004vj}, appears to be the magnetic part of the
symplectic invariant scalar potentials containing, besides the
magnetic charges $m^{I\L}$ $I=1,2$, also the Abelian electric
charges $e^I_\L$. This raises the question whether we can further
modify the shifts in the fermionic transformation laws in such a
way that the resulting potentials contain both electric and
magnetic charges in a symplectic invariant setting.\\
We now show that it is actually possible to add further
Fayet--Iliopoulos "electric" mass--shifts to the fermion
transformation laws which contain electric charges, in such a way
that the total shifts are symplectic invariant, thereby giving
rise to a symplectic invariant scalar potential. The technical
procedure to arrive to such an extension is explained in detail in
the Appendix B. There we show that, using Bianchi identities in
superspace, the global Fayet--Iliopoulos term appearing in the
transformation laws of the fermions can be written in the
following symplectic invariant way:

\begin{eqnarray}
S_{AB}\equiv S_{AB}^{(e)}+S_{AB}^{(m)}& =& \imez \s^x_{AB} \,
\o^x_I \sx( e^I_\L L^\L - m^{I\L} M_\L \dx) \label{shiftSem}\\
W^{iAB}\equiv W^{iAB(e)}+W^{iAB(m)} & = & {\rm i} g^{i\bar \jmath}
\s^{x\,AB} \o^x_I \sx(e^I_\L \ol{f}_{\bar\jmath}^\L -
m^{I\L} \ol{h}_{\bar\jmath\L}\dx)\label{shiftWem}\\
N^{A}_\a\equiv N^{A(e)}_\a +N^{A(m)}_\a &=& 2\, \op{U}_I{}^A{}_\a
\sx( e^I_\L L^\L - m^{I\L} M_\L \dx) \label{shiftNem}
\end{eqnarray}

\noi where we have assumed that the vector ($m^{I\L}, e^I_\L$)
transforms as the symplectic section $V$ of equation
(\ref{sympsect}). In this case the scalar potential turns out to
be symplectic invariant and its explicit form is given by:

\begin{eqnarray}
\label{eq:pot} \op{V}&=& 4 \sx( \op{M}_{IJ} - \o^x_I \o^x_J \dx)
\sx( m^{I\L} \ol{M}_\L -e^I_\L \ol{L}^\L \dx) \sx( m^{J\S}M_\S
-e^J_\S L^\S \dx) + \nn\\
&&+ \o_I^x \o^x_J\sx( m^{I\L} , e^I_{\L} \dx) \op{S}
\begin{pmatrix}{m^{J\S} \cr e^J_{\S}}\end{pmatrix}
\end{eqnarray}

\noi where the matrix $\op{S}$ is a symplectic matrix given
explicitly by:

\begin{equation}\label{esse}
\op{S} = -\mez \begin{pmatrix}{ \Im_{\L\S} + \sx( \Re \Im^{-1} \Re
\dx)_{\L\S} & -\sx( \Re \Im^{-1} \dx)_{\L}{}^{\S}\cr
-\sx(\Im^{-1}\Re\dx)^{\L}{}_{\S} & \Im^{-1|\L\S}}\end{pmatrix}.
\end{equation}

\noi $\Im_{\L\S}$ and $\Re_{\L\S}$ being the imaginary and real
part of the period matrix $\op{N}_{\L\S}$ of the special geometry.
Furthermore the electric and magnetic charges must satisfy the
constraint:

\begin{equation}\label{constr}
    e^I_\L m^{J\L} - e^J_\L m^{I\L} = 0
\end{equation}

\noi which follows from the Ward identity of supersymmetry.\\
As before the dependence of the scalar potential on the $q^u$'s
appears only through the SU$(2)\otimes$SO$(n_T)$ vector
$\o_I^x(q^u)$, while the dependence on $z$, $\ol z$ appears
through the symplectic matrix of equation (\ref{eq:pot}) and the
covariantly holomorphic sections $L^\L$, $M_\L$ of the special
geometry.\\
>From the point of view of dualization of $N=2$ standard theory the
electric charges $e^I_\L$ can be associated to the gauging of the
translational group of the axionic symmetries with generators
$T_\L$ and subsequent dualization of the axions, while the
magnetic charges $m^{\L I}$ can be thought to be associated with
the isometry algebra of the same translation group, with
generators $T^\L$, in the dual theory. Indeed if we would start in
the standard $N=2$ supergravity from a purely magnetic theory with
magnetic vector potentials $A_\L$, then the Killing vectors
associated to the translational group of the axions would
naturally have an upper index $\L$ as the generators of the
translation group of the dual
theory.\\
In the case of Calabi--Yau compactification from Type IIB string
theory the scalar potential (\ref{eq:pot}) simplifies due to the
important cancellation\break $\op{M}_{IJ} - \o^x_I \o^x_J = 0$
\cite{Dall'Agata:2001zh}. Moreover \cite{Taylor:1999ii},
\cite{Dall'Agata:2001zh}, \cite{Louis:2002ny} one finds that the
tensors $\o_I^x$ are given by \cite{Ferrara:ik}:

\begin{eqnarray}
& \o_1^{(1)}=0;\quad \o_1^{(2)}=0;\quad\o_1^{(3)}=e^\varphi&\nn\\[2mm]
& \o_2^{(1)}=-e^\varphi {\rm Im}\t;\quad
\o_2^{(2)}=0;\quad\o_2^{(3)}=e^\varphi {\rm Re}\t.&\label{oconn}
\end{eqnarray}

\noi where $\varphi$ is the quaternionic scalar field defined as
$\tilde K$ in reference \cite{Ferrara:ik}. Using the definition
\eq{oconn} one easily sees that the scalar potential for
Calabi--Yau compactifications can be rewritten as:

\begin{equation}
\op{V}_{CY}=-\frac{1}{2}e^{2\varphi} \sx[ \sx( e_\L - \ol
\op{N}_{\L\S} m^\S \dx) \sx(\im\dx)^{-1|\L\G} \sx( \ol{e}_\G -
\op{N}_{\G\D} \ol{m}^\D \dx) \dx]
\end{equation}

\noi where

\begin{equation}
e_\L\equiv e^1_\L+\t e^2_\L;\quad\quad m^\L\equiv m^{1\L}+\t
m^{2\L},
\end{equation}

\noi $\tau$ being the ten dimensional complex dilaton. This is
exactly the potential appearing in the quoted references, where
the constraint (\ref{constr}) reduces to the "tadpole cancellation
condition" \be e^1_\L m^{2\L} - e^2_\L m^{1\L} =0\ee We note that
the constraint (\ref{constr}) (generalized tadpole cancellation
condition) is now a consequence of the supersymmetry Ward identity
\eq{ward} that one uses to
compute the scalar potential \cite{Sommovigo:2004vj}.\\
If one thinks of this theory as coming from dualization of the
axionic coordinates $q^I$ of the quaternionic manifold of the
standard $N=2$ supergravity, the electric charges would appear as
due to the gauging of the translational group of isometries $q^I
\to q^I + c^I$. Indeed the quaternionic prepotential appearing in
the gauge shifts of the fermions would be given in this case by
the formula:

\begin{equation}\label{prep}
    \op{P}^x_\L = \o^x_I e^I_\L
\end{equation}

\noi $e^I_\L$ being the constant Killing vectors of the
translations \cite{Andrianopoli:1996ve}. \\
Note that the procedure adopted in reference
\cite{Dall'Agata:2003yr} does not seem suitable in order to find
the Abelian electric charges $e^I_\L$. In fact, it is evident
that, even though formally they give rise to the same kind of
structures \cite{Dall'Agata:2003yr}, the Abelian electric charges
$e^I_\L$ which pair together with the masses (or magnetic charges)
$m^{I\L}$ to reconstruct symplectic invariant structures
\cite{Sommovigo:2004vj}, \cite{Taylor:1999ii},
\cite{Louis:2002ny}, can not descend from the gauging of Abelian
isometries of $\op{M}_T$ since after dualization such axionic
symmetries are not present on the residual $\s$--model on
$\op{M}_T$. One can understand it by considering the relation with
standard $N=2$ gauged supergravity \cite{Sommovigo:2004vj} or just
observing that if the Abelian electric charges must combine into a
symplectic vector with the mass parameters

\begin{equation}
\op{K}^I=\begin{pmatrix}{m^{I\L}\cr e^I_\L}\end{pmatrix}
\end{equation}

\noi they must carry the same index $I$ of the tensors, that is of
the axions which have been dualized. Even if we suppose that not
all the axions have been dualized and thus there are some
translational isometries  left on the reduced manifold, the
corresponding Killing vectors would not carry the index
$I=1,\dots,n_T$.

\section{The Lagrangian}
\label{Lagrangian} \setcounter{equation}{0}

In this section we present the Lagrangian and the supersymmetry
transformation laws under which it is invariant. The method we
used to derive the Lagrangian is the so called geometrical or
rheonomic approach which is greatly facilitated when the solution
of Bianchi identities in superspace is already known. The
superspace geometrical Lagrangian and the solution of the Bianchi
identities in superspace are shortly discussed in Appendix B. We
limit ourselves in the main text to give the Lagrangian and the
supersymmetry transformation laws up to 4--fermions and
3--fermions respectively, while the completion of the formulae
containing quadrilinear and trilinear fermions are given in
Appendix A. Our result is the following\footnote{For the notations
of $N=2$ theory and special geometry we refer the reader to the
standard papers \cite{Andrianopoli:1996ve},
\cite{Andrianopoli:1996cm}}

\begin{equation}
S= \int {\rm d}^4 x \sqrt{-g} \, \mathcal{L}
\end{equation}

\begin{equation}\label{laglag}\op{L} = \op{L}_\bos + \op{L}_{\rm ferm}^{\rm
kin} + \op{L}_{\rm Pauli} + \op{L}_{\rm shifts} + \op{L}_{\rm 4f}
\end{equation}

\begin{eqnarray}
  \label{eq:lagbos}
  \op{L}_\bos &=& -\mez \op{R} + g_{i\bar \jmath} \, \der^\m z^i
  \der_\m {\ol z}^{\bar \jmath}+ {\rm i} \sx( \ol\op{N}_{\L\S}
  \hat\op{F}^{-\L}_{\m\n}\hat\op{F}^{-\L\m\n} - \op{N}_{\L\S}
  \hat\op{F}^{+\L}_{\m\n}\hat\op{F}^{+\L\m\n}\dx) \nn\\
  && + 6 \op{M}^{IJ} \op{H}_{I\m\n\r} \op{H}_J{}^{\m\n\r} +g_{uv} \der^\m q^u
  \der_\m q^v+2 h^\m_I A^I_u \na_\m q^u + \nn \\
  && -2 e^I_\L \sx(\hat\op{F}^\L_{\m\n} - m^{J\L}
  B_{J\m\n}\dx) B_{I\r\s} \frac{\e^{\m\n\r\s}}{\sqrt{-g}}
  -\op{V}(z, \ol z, q^u)\\
  \op{L}_{\rm ferm}^{\rm kin} &=& \frac{\e^{\m\n\l\s}}{\sqrt{-g}} \sx( \ol \psi^A_\m \g_\n \r_{A|\l\s} -
  \ol\psi_{A|\m}\g_\n\r^A_{\l\s}\dx) -\imez g_{i \bar \jmath}\sx(
  \ol \l^{i A} \g^\m \na_\m \l^{\bar \jmath}_A \dx.+\nn\\
  && \sx.+\ol \l^{\bar \jmath}_A \g^\m \na_\m \l^{i A} \dx) -{\rm i}\sx(
  \ol \z^\a \g^\m \na_\m \z_\a +\ol \z_\a \g^\m
  \na_\m\z^\a\dx)\label{eq:lagkinf}\\
  \op{L}_{\rm Pauli} &=& -g_{i\bar \jmath}\sx( \der^\m z^i\ol \psi^A_\m
  \l^{\bar \jmath}_A + \der_\m z^i \ol \psi^A_\n \g^{\m\n} \l^{\bar \jmath}_A
  + \cc \dx)\nn\\
  &&-2\sx( P_{u\,A\a} \der^\m q^u \ol\psi^A_\m \z^\a + P_{u\,A\a} \der_\m q^u
  \ol\psi^A_\n \g^{\m\n} \z^\a +\cc \dx) \nn\\
  && + 6 {\rm i} \op{M}^{IJ} \op{H}_I{}^{\m\n\r} \sx[ {\cal U}_J{}^{A\a}\ol
  \psi_{A|\m} \g_{\n\r} \z_\a - {\cal U}_{J\, A\a} \ol\psi^A_\m \g_{\n\r}
  \z^\a \dx]+\nn\\
  && - 24 {\rm i} \op{M}^{IJ} h_I{}^\m \D_J{}^\a{}_\b \ol \z_\a \g_\m
  \z_\b + 24 {\rm i} A^I_u \D_I{}^\a{}_\b \ol \z_\a \g^\m \z_\b
  \na_\m q^u + \nn\\
  &&+\sx\{\hat\op{F}_{\m\n}^{-\L} \sx( {\rm Im}\op{N}\dx)_{\L\S}
  \sx[ 4L^\S \sx(\ol\psi^{A|\m}\psi^{B|\n}\dx)^- \e_{AB} +4{\rm i}
  f_{\bar \imath}^\S \sx(\ol \l^{\ol \imath}_A \g^\m \psi_B^\n \dx)^-
  \e^{AB}\dx.\dx.\nn\\
  &&\sx.\sx. +\mez \na_i f_j^\S \ol\l^{iA} \g^{\m\n} \l^{jB} \e_{AB} - L^\S \ol
  \z_\a \g^{\m\n}\z_\b \IC^{\a\b}\dx] +\cc \dx\}+\nn\\
  &&+2 \op{M}^{IJ} h_I^\m \sx( {\cal U}_J{}^{A\a} \ol\psi_{A|\m}\z_\a +
  {\cal U}_{J\,A\a} \ol\psi^A_\m \z^\a \dx)+\nn\\
  &&+2{\rm i} \op{M}^{IJ} h_I^\m \D_{J\,\a}{}^\b\z_\b\g_\m\z^\a
  \label{eq:lagpauli}\\
  \op{L}_{\rm shifts}&=& \sx(2 \ol S^{AB} \ol\psi_A^\m \g_{\m\n}
  \psi^\n_B + {\rm i} g_{i\bar \jmath} W^{iAB} \ol\l^{\bar \jmath}_A \g_\m
  \psi^\m_B  +2{\rm i} N^A_\a \ol\z^\a \g_\m \psi^\m_A\dx.  + \nn\\
  &&+ \mez \na_u N^A_\a P^u{}_{A\b}\ol\z^\a\z^\b +2 \na_{\bar \imath}
  N^A_\a \ol \z^\a \l^{\bar \imath}_A + \nn\\
  &&+\sx. \frac{1}{3}g_{i\bar \jmath} \na_k W^{\bar \jmath}_{AB} \ol \l^{iA}
  \l^{kB}\dx)+\cc +\op{L}_{\rm 4f}^{\rm non\, inv} + \op{L}_{\rm 4f}^{\rm inv}
  \label{eq:lagshifts}
\end{eqnarray}

\noi where the scalar potential

\begin{eqnarray}
\label{eq:pote} \op{V}&=& 4 \sx( \op{M}_{IJ} - \o^x_I \o^x_J \dx)
\sx( m^{I\L} \ol{M}_\L -e^I_\L \ol{L}^\L \dx) \sx( m^{J\S}M_\S
-e^J_\S L^\S \dx) + \nn\\
&&+ \o_I^x \o^x_J\sx( m^{I\L} , e^I_{\L} \dx) \op{S}
\begin{pmatrix}{m^{J\S} \cr e^J_{\S}}\end{pmatrix}
\end{eqnarray}

\noi coincides with the expression given in equation
(\ref{eq:pot}). We have rewritten it in the Lagrangian for
completeness. Notations are as follows (for what is concerned the
special geometry quantities we use the standard notations of
reference \cite{Andrianopoli:1996ve}, in particular $i,j$, $\bar
\imath, \bar \jmath$ are curved indices on the special K\"ahler
manifold): covariant derivatives on the fermions are defined in
terms of the 1--form SU$(2)$--connection $\o^{AB}$, the K\"ahler
U$(1)$ connection, $Q$, the Sp$(2n_H)$ connection $\D^{\a\b}$ and
Christoffel connection $\G^i{}_j$ defined on $\op{M}_{SK} \otimes
\op{M}_T$ (covariant derivatives are defined in Appendix B). Note
that $f^\S_j$ being part of the symplectic section of special
geometry has a derivative covariant with respect to the K\"ahler
connection. Furthermore $P_{uA\a}$ is a "rectangular vielbein"
\cite{Dall'Agata:2003yr} related to the metric $g_{uv}$ of the
$\s$--model by the relation $P_u{}^{A\a}P_{vA\a} = g_{uv}$, and
$P^{uA\a}= g^{uv}P_v{}^{A\a} $. \\
Besides the field--strengths of the tensors, all the $I,J$ indexed
quantities, namely $A^I_u$, $\op{U}_I{}^{A\a}$, $\o_{IA}{}^B$,
$\D_{I\a}{}^\b$ are vectors on an SO$(n_T)$ bundle defined on the
$\s$--model, whose metric is $\op{M}_{IJ}$. We note that, if one
thinks of this Lagrangian as coming from the $N=2$ standard
supergravity \cite{Andrianopoli:1996ve}, all these $I$--indexed
quantities can be interpreted as the remnants of the original
quaternionic metric $h_{\hat u \hat v}$ ($M_{IJ}$, $A^I_u$),
vielbein $\op{U}^{A\a}_{\hat u}$ ($\op{U}_I{}^{A\a}$), SU$(2)$ and
Sp$(2n)$ 1--form connections $\o^{AB}$ and $\D^{\a\b}$ after
dualization of the $q^I$ coordinates ($q^{\hat u} = (q^u, q^I)$)
of the quaternionic manifold. Moreover we have defined:

\begin{eqnarray}
\op{F}_{\m\n}^\L &=& \partial_{[\m} A^\L_{\n]} \\
\op{F}^{\pm\L}_{\m\n} &=& \frac{1}{2} \sx(\op{F}^\L_{\m\n} \pm
\imez \e_{\m\n\r\s} \op{F}^{\L\r\s}\dx)\\
\op{H}_{I\m\n\r} &=& \partial_{[\m} B_{I\,\n\r]}\\
h_{I\m} &=& \e_{\m\n\r\s} \op{H}_I{}^{\n\r\s}.
\end{eqnarray}

\noi The Lagrangian \eq{laglag} is invariant under the following
supersymmetry transformations:

\begin{eqnarray}
\d\psi_{A|\m} &=& \op{D}_\m \e_A -h^I_\m \o_{I\, A}{}^B \ve_B
+\nn\\
&&\quad + \sx[ {\rm i} S_{AB}\eta_{\m\n}+e_{AB} T^-_{\m\n}\dx]
\g^\n\e^B  + {\rm 3\, fermions}\label{trasfgrav0}\\
\d\l^{iA}&=& {\rm i} \sx(\na_\m z^i \dx)\g^\m \e^A +G^{-i}_{\m\n}
\g^{\m\n} \e_B \e^{AB}+ W^{iAB}\e_B+ {\rm 3\, fermions}
\label{gaugintrasfm0}\\
\d\z_\a&=&{\rm i}P_{uA\a} \der_\m q^u \g^\m \ve^A- {\rm i} h^I_\m
\op{U}_{IA\a} \g^\m \ve^A + N_\a^A
\ve_A + {\rm 3\, fermions}\label{iperintrasf0}\\
\d V^a_\m &=& -{\rm i} \ol \psi_{A \m} \g^a \e^A -{\rm i}\ol
\psi^A_\m \g^a
\e_A\label{vieltrasf0}\\
\d A^\L_\m &=& 2 L^\L \ol \psi^A_\m \e^B \e_{AB} + 2 \ol L^\L \ol
\psi_{A\m}
\e_B \e^{AB} +\nn\\
&&+\sx({\rm i} f^\L_i \ol \l^{iA} \g_\m \e^B \e_{AB} +{\rm i} \ol
f^\L_{\bar\imath} \ol\l^{\bar\imath}_A \g_\m \e_B \e^{AB}\dx)
\label{gaugtrasf0}\\
\d B_{I\m\n} &=& - \imez \sx(\ol \ve_A \g_{\m\n} \z_\a
\op{U}_I{}^{A\a} -
\ol \ve^A \g_{\m\n} \z^\a \op{U}_{IA\alpha}\dx)+\nn\\
&& -\o_{I C}{}^{A} \sx( \ol\ve_A \g_{[\m}\psi^C_{\n]}+
\ol\psi_{[\m A}
\g_{\n]} \ve^C\dx) \label{trasfb}\\
\d z^i &=& \ol \l^{iA}\e_A \label{ztrasf0}\\
\d \ol z^{\bar\imath}&=& \ol\l^{\bar\imath}_A \e^A \label{ztrasfb0}\\
\d q^u &=& P^u{}_{A\a} \sx(\ol \z^\a \e^A + \IC^{\a\b}\e^{AB}\ol
\z_\b \e_B \dx)\label{qtrasf0}
\end{eqnarray}

\noi where the dressed field--strengths appearing in the
transformation laws of the gravitino and gaugino fields are
defined as:

\begin{eqnarray}
T^-_{\m\n} &=& 2{\rm i}\sx(\im \dx)_{\L\S} L^\S \hat
\op{F}_{\m\n}^{\L -} + {\rm \,bilinear\, fermions}\label{T-def}\\
G^{i-}_{\m\n} &=& - g^{i\bar\jmath}\, \ol f^\G_{\bar\jmath}
\sx(\im \dx)_{\G\L} {\hat\op{F}}^{\L -}_{\m\n} + {\rm \,bilinear\,
fermions} \label{G-def}
\end{eqnarray}

\noi where the bilinear fermions are the same as in standard $N=2$
supergravity and are given explicitly in the Appendix
\ref{app:compl}. We have written the transformation laws only for
the chiral spinor fields ($\psi_{A\m}$, $\l^{iA}$, $\z_\a$); the
transformation laws for the anti--chiral fields ($\psi^A_\m$,
$\l^{\bar \imath}_A$,
$\z^\a$) are immediately obtained from the chiral ones.\\
Finally, the shift matrices $S_{AB}$, $W^{i\, AB}$ and $N^A_\a$
are given by equations (\ref{shiftSem},\\
\ref{shiftWem},\ref{shiftNem}). We also note that the given
Lagrangian is invariant under the gauge transformation

\begin{eqnarray}
\label{eq:gauge} \d B_I &=& {\rm d} \L_I \nn\\
\d A^\L &=& -2 m^{I\L} \L_I \nn
\end{eqnarray}

\noi where $\L_I$ is a $1$--form, if and only if the following
generalized tadpole cancellation condition is satisfied

\begin{equation}
e^I_\L m^{J\L}-e^J_\L m^{I\L}=0.
\end{equation}

\noi Indeed $\hat \op{F}^\L$ is invariant by itself while the
topological term is invariant only if equation (\ref{constr}) is
satisfied. As observed in the previous section, the same
consistency condition also appears from the supersymmetry Ward
identity from which one can compute the scalar potential.\\
As far as the symplectic invariance is concerned
\cite{Gaillard:1981rj}, nothing is changed with respect to the
standard $N=2$ supergravity. Indeed if we consider the bosonic
part of the Lagrangian for the vector and tensor fields, namely:

\begin{eqnarray}
\label{kintop}
  \op{L}^\bos_{\rm vect} +\op{L}^\bos_{\rm top} &=& {\rm i}
  \sx( \ol\op{N}_{\L\S}\hat\op{F}^{\L-}_{\m\n}\hat \op{F}^{\S-\m\n}-
  \op{N}_{\L\S} \hat\op{F}^{\L+}_{\m\n} \hat\op{F}^{\S +\m \n}
  \dx)\nn\\
  &&-4{\rm i}e^I_\L \sx(\hat \op{F}^{\L-}_{\m\n} B^{-\m\n}_I -
  \hat \op{F}^{\L+}_{\m\n} B^{+\m\n}_I\dx) +\nn\\
  &&+ 4 {\rm i} m^{J\L} e^I_\L B_J^{-\m\n} B_{I\m\n}^- - 4
  {\rm i} m^{J\L} e^I_\L B_J^{+\m\n} B_{I\m\n}^+
\end{eqnarray}

\noi and define

\begin{equation}
\op{G}_{\L\m\n}^{(\bos)}{}^\mp = \mp \imez \frac{\d
\op{L}^{(\bos)}}{\d \hat \op{F}^{\L\mp\m\n}} = \ol{\op{N}}_{\L\S}
\hat \op{F}^{\S\mp}_{\m\n} -2 e^I_\L B_{I\m\n}^\mp \label{gbos}
\end{equation}

\noi then equation (\ref{kintop}) can be rewritten as

\begin{eqnarray}
\label{kintop2} &{\rm i}
\sx(\hat\op{F}^{\L-}_{\m\n}\op{G}_\L^{(\bos)\, - \m\n} - 2 e^I_\L
B^-_{I\m\n} \hat \op{F}^{\L-\m\n}+4 e^I_\L m^{J\L} B_{I\m\n}^-
B_J^{-\m\n}\dx) + {\rm c.c.} =&\nn\\
& ={\rm i} \sx[ \op{F}^{\L-} \op{G}_\L^{(\bos)\,-} + 2 B_I^- \sx(
m^{I\L} \op{G}^-_\L - e^I_\L \op{F}^{\L-} \dx) \dx] + \cc&
\end{eqnarray}

\noi The first term in (\ref{kintop2}) is the usual vector kinetic
term of the standard $N=2$ theory while the additional term is a
symplectic invariant. Indeed one can easily verify that
$\op{G}_\L^{(\bos)\,-}$, as defined in (\ref{gbos}), transforms as
the lower component of the symplectic vector ($\op{F}^{\L -}$,
$\op{G}_\L^{(\bos)\,-}$), provided $\op{N}_{\L\S}$ transforms
under the symplectic transformation (\ref{sympmat}) as usual,
namely:

\begin{equation}
\op{N}'_{\L\S} = \sx(C + D\op{N} \dx)\sx(A + B\op{N} \dx)^{-1}
\end{equation}

\noi On the other hand the vector ($m^{I\L}$, $e^I_\L$) was
already supposed to transform in the same way as the vector
($L^\L$, $M_\L$) as does the symplectic vector ($\op{F}^{\L -}$,
$\op{G}^{\L-}$) of the special geometry.\\
If we also take into account the fermionic sector of the
Lagrangian then $\op{G}^{(\bos)}$ is completed with bilinear
fermions:

\begin{eqnarray}
\op{G}_{\L\m\n}^- &=& \op{G}_{\L\m\n}^{(\bos)}{}^- -\imez
(\im)_{\L\S} \op{D}^{\S-}_{\m\n} \nn\\
\op{G}_{\L\m\n}^+ &=& \sx(\op{G}_{\L\m\n}^- \dx)^*
\end{eqnarray}

\noi where $\op{D}^{\L-}_{\m\n}$ is the coefficient of
$\op{F}^{\S-}_{\m\n} {\rm Im}\op{N}_{\L\S}$ in $\op{L}_{\rm
Pauli}$. Following the argument of reference
\cite{Ceresole:1995jg}, the full Lagrangian is on--shell
symplectic invariant (or invariant up to the $\op{F}\op{G}$ term)
by adding non--symplectic invariant 4--fermion terms $\op{L}_{\rm
4f}^{\rm non\, inv}$. This non--invariant 4--fermion terms are
exactly the same as in the standard $N=2$ theory because so is the
term $- \imez (\im)_{\L\S} \op{D}^{\S-}_{\m\n}$ and their explicit
form is given in the Appendix A. $\op{L}^{\rm inv}_{4 {\rm f}}$ is
instead fixed only by supersymmetry. Its explicit form again
coincides with that given in the standard theory except for some
additional terms as explained in Appendix A.\\
We also note that the equations of motion and Bianchi identities
for the vectors have exactly the same form as in the standard
theory, namely:

\begin{eqnarray}
\der_\m {\rm Im} \op{F}^{\L-\,\m\n} &=& 0 \label{eq:eomf}\\
\der_\m {\rm Im} \op{G}^-_\L{}^{\m\n} &=& 0 \label{eq:bianchif}
\end{eqnarray}

\noi where $\op{G}^-_{\L\m\n}$ differs from the analogous one of
the standard $N=2$ theory by the term linear in $B_{I\m\n}$
appearing in the equation (\ref{gbos}), and they are symplectic
covariant.\\
Finally we note that the generalized tadpole cancellation
condition (\ref{constr}) implies that the symplectic vectors
$m^{I\L}, e^I_\L$ are all parallel. Therefore by a symplectic
rotation we can choose the gauge where all the magnetic charges
$m^{I\L}$ are zero. In this case we are in a strictly perturbative
regime. In an analogous way we could also rotate the charge vector
in such a way that all the electric charges $e^I_\L$ are zero and
we may think of this regime as a perturbative regime for a purely
magnetic theory.

\subsection{Adding a semisimple gauging}
\label{adding}

It is now immediate to generalize the theory given in the previous
section by gauging the group of the isometries of the $\s$--model
parametrized by the scalars $q^u$ of the scalar tensor multiplet.
Let us call $\op{G}$ the group of isometries that can be gauged on
such a manifold. The gauging due to $\op{G}$ was given in a
general form in reference \cite{Dall'Agata:2003yr} where however
there was no consideration of the Fayet--Iliopoulos terms giving
rise to the electric charges. Such a gauging breaks of course the
symplectic invariance of the equations of motion since it involves
only the gauging associated to electric vector potentials. In
order to distinguish between the Fayet--Iliopoulos terms and the
new terms associated to the gauging we split the index $\L$ into
($\L = \hat\L$, $\check \L$), where $\hat\L = 1, \dots ,n_T$ is
used for the electric and magnetic charges and the related Special
Geometry sections associated to the deformations coming from
Fayet--Iliopoulos terms, while we denote by $\check \L$, the index
running on the adjoint representation of the group $\op{G}$,
associated to the electric vector potentials $A^{\check \L}_\m$
that are gauged. Referring to reference \cite{Dall'Agata:2003yr},
we know that the fermionic shifts acquire a further term so that
the total shifts are now given by

\begin{eqnarray}
S_{AB} & =& \imez\,\s^x_{AB} \, g \, \op{P}^x_{\check\L}
L^{\check\L} + \imez \, \s^x_{AB} \, \o_I^x \sx( e^I_{\hat\L}
L^{\hat\L} - m^{I\hat\L} M_{\hat\L} \dx) \label{shiftS3}\\
N^\a_A & = & -2 \, g \,\sx( P_u{}_A{}^\a + A^I_u \op{U}_I{}_A{}^\a
\dx) k^u_{\check\L} L^{\check\L} -2 \, \op{U}_I{}_A{}^\a \sx(
e^I_{\hat\L} L^{\hat\L} - m^{I\hat\L} M_{\hat\L} \dx) \label{shiftN3}\\
W^{iAB} & = & -{\rm i}\,g\, \e^{AB} g^{i\bar \jmath} P_{\check\L}
f^{\check\L}_{\bar \jmath} + {\rm i} \, g^{i\bar \jmath}
\s^{x\,AB} g {\cal P}^x_{\check\L}\ol f_{\bar \jmath}^{\check\L} + \nn\\
&&+ {\rm i}\,g^{i\bar\jmath} \s^{x\,AB} \o^x_I \sx(e^I_{\hat\L}\ol
f_{\bar\jmath}^{\hat\L} - m^{I\hat\L} h_{\hat\L \bar\jmath} \dx)
\label{shiftW3}
\end{eqnarray}

\noi which in fact reduce to those given in reference
\cite{Dall'Agata:2003yr} if we set $e^I_{\hat \L}=0$. Here
$P_{\check \L}$ is the prepotential of the special geometry,
$\op{P}^x_{\hat \L}$ is a new quantity which from the point of
view of dualization of standard $N=2$ supergravity, can be thought
as the quaternionic prepotential restricted to $\op{M}_T$ and
$k^u_{\check\L}$ is the Killing vector of the gauged isometries.
As we pointed out before, from the point of view of dualizing the
standard $N=2$ theory with quaternions the terms in $e^I_{\hat
\L}$ appearing in the shifts can be thought of as coming from a
translational gauging of the axionic symmetries on the
quaternionic manifold before dualization so that, from this point
of view, the gauge group which is gauged is $\op{T} \otimes
\op{G}$ where $\op{T}$ is an Abelian translation group gauging the
axionic symmetries.\\
The Lagrangian for the gauged theory can be written formally in
exactly the same way as the Lagrangian (\ref{laglag}) provided we
make the following modification (for more details see
\cite{Andrianopoli:1996ve}): the covariant derivatives on the
fermions are now defined in terms of the gauged connections $\hat
\o^{AB}$, $\hat \D^{\a\b}$, $\G^i{}_j$ defined in reference
\cite{Andrianopoli:1996ve} while the ordinary derivatives acting
on the scalars $z^i$ and $q^u$ are now replaced by gauge
derivatives:

\begin{eqnarray}
\der_\m q^u &\to& \na_\m q^u \equiv \der_\m q^u + k^u_{\hat \L}
A^{\hat \L}_\m \label{gaugqder}\\
\der_\m z^i &\to& \na_\m z^i \equiv \der_\m z^i + k^i_{\hat \L}
A^{\hat \L}_\m \label{gaugzder}\\
\der_\m \ol z^{\bar \imath} &\to& \na_\m \ol z^{\bar \imath}
\equiv \der_\m \ol z^{\bar \imath} + k^{\bar \imath}_{\hat \L}
A^{\hat \L}_\m \label{gaugzbarder}
\end{eqnarray}

\noi Furthermore the scalar potential takes additional terms which
are discussed in the next section. In order not to have too messy
formula we do not perform the splitting $\L \to \hat\L$, $\check
\L$ on such generalized Lagrangian. However one must pay attention
to the fact that saturation of indices $\L$, $\S$ in the
Lagrangian have to be done according to this splitting; in
particular saturation of the $\L$ indices with electric and
magnetic charges run only on $\hat \L$ and furthermore we have of
course

\begin{eqnarray}
\hat F^{\check\L} &=& {\rm d} A^{\check\L} + \mez
f_{\check\S\check\Pi}{}^{\check\L} A^{\check\S}A^{\check\Pi}\\
\hat F^{\hat \L} &=& {\rm d} A^{\hat\L} + m^{I\hat \L} B_I.
\end{eqnarray}

\noi However the splitting $\L \to \hat \L$, $\check \L$ will be
performed explicitly in the next section for the discussion of the
scalar potential in the presence of gauging.

\subsection{The scalar potential for the gauged theory}
\label{discpot}

The explicit form of the scalar potential is:

\begin{eqnarray}\label{poten}
\op{V}&=& g_{i\bar\jmath} k^i_{\check \L} k^{\bar\jmath}_{\check
\S} \ol L^{\check\L} L^{\check\S} +4
\sx(m^{I\hat\L},e^I_{\hat\L},k^u_{\check\L}\dx) \op{Q}
\begin{pmatrix}{m^{J\hat\S}\cr e^J_{\hat\S}\cr
k^v_{\check\S}}\end{pmatrix}+
\nn\\
&&+\sx(m^{I\hat\L} \o_I^x,e^I_{\hat\L}\o^x_I,{\cal P}^x_{\check\L}
\dx) \sx(\op{R} + \op{S} \dx)
\begin{pmatrix}{m^{J\hat\S}\o_J^x \cr e^J_{\hat\S} \o^x_J \cr
{\cal P}^x_{\check\S}}\end{pmatrix}
\end{eqnarray}

\noi where we have defined the following matrices

\begin{eqnarray}
\op{Q} &=& \begin{pmatrix}{\op{M}_{IJ} \ol M_{\hat\L} M_{\hat\S} &
-\mathcal{M}_{IJ} \ol M_{\hat\L} L^{\hat\S} & -A_v^K \op{M}_{IK}
\ol M_{\hat\L} L^{\check\S} \cr -\mathcal{M}_{IJ} \ol L^{\hat\L}
M_{\hat \S} &
 \mathcal{M}_{IJ}\ol L^{\hat\L} L^{\hat\S} & A_v^K \op{M}_{KI}\ol L^{\hat\L}
 L^{\check\S} \cr -A_u^K \op{M}_{KJ} \ol L^{\check\L} M_{\hat\S}
& A^K_u \op{M}_{KJ} \ol M_{\hat\S} L^{\check\L}& \sx( g_{uv}
+A^I_u A^J_v \mathcal{M}_{IJ}\dx) \ol L^{\check\L}
L^{\check\S}}\end{pmatrix} \nn\\
\op{R}&=& -4 \begin{pmatrix}{ \ol M_{\hat\L} M_{\hat\S} & -\ol
M_{\hat\L} L^{\hat\S}& -\ol L^{\check\S} M_{\hat\L} \cr -\ol
L^{\hat\L} M_{\hat\S} & \ol L^{\hat\L} L^{\hat\S} & \ol L^{\hat\L}
L^{\check\S} \cr - L^{\check\L} \ol M_{\hat\S}& \ol L^{\check\L}
L^{\hat\S}&\ol L^{\check\L}
L^{\check\S}}\end{pmatrix}\nn\\
\op{S}&=& -\mez \begin{pmatrix}{\Im_{\hat\L\hat\S} +
\sx(\Re\Im^{-1}\Re\dx)_{\hat\L\hat\S}&
-\sx(\Re\Im^{-1}\dx)_{\hat\L}{}^{\hat\S}& -\sx(\Im^{-1}
\Re\dx)_{\hat\L}{}^{\check\S} \cr
-\sx(\Im^{-1}\Re\dx)^{\hat\L}{}_{\hat\S}&\Im^{-1|\hat\L\hat\S}&
\Im^{-1|\hat\L\check\S}\cr -\sx(\Re
\Im^{-1}\dx)^{\check\L}{}_{\hat\S}
&\Im^{-1|\check\L\hat\S}&\Im^{-1|\check\L\check\S}}\end{pmatrix},
\end{eqnarray}

\noi and where we have set as before $\Im_{\L\S} = \im_{\L\S}$,
$\Re_{\L\S} = \re_{\L\S}$.\\
The derivation of the scalar potential has been found from the
supersymmetry Ward identity:

\begin{equation}
\label{ward} \delta_B^A {\cal V}  \,= \,-12\,\overline{S}^{CA} \,
S_{CB}\,+\,g_{i\bar \jmath}W^{iCA}\,W_{CB}^{\bar \jmath}\,
+\,2N^A_\alpha \, N^\alpha_B
\end{equation}

\noi This identity implies that all the terms proportional to a
Pauli $\s$--matrix must cancel against each other. We find besides
the constraint (\ref{constr}) the following new constraint:\\

\begin{equation}
f_{\check\Lambda\check\Sigma}{}^{\check\Delta} \,
(\im)_{\check\Delta\hat\Pi} \, m^{I\hat\Pi} =
f_{\check\Lambda\check\Sigma}{}^{\check\Delta}
\tilde{m}^{I}_{\check\Delta}= 0\,, \label{eq:contr}
\end{equation}

\noi where

\begin{equation}
\tilde{m}^{I}_{\check\Delta}\equiv (\im)_{\check\Delta\hat\Pi} \,
m^{I\hat\Pi} \,.
\end{equation}

\noi This condition was already discussed in reference
\cite{Dall'Agata:2003yr}. There it was noted that this equation
means that $\tilde{m}^{I}_{\check\Delta}$ are the coordinates of a
Lie algebra element of the gauge group commuting with all the
generators, in other words a non trivial element of the center
${\mathbb Z}(\op{G})$. From the point of view of dualization of
the standard $N=2$ supergravity with gauged translations this
condition has a clear interpretation. Indeed since our theory can
be thought of as coming from the gauging of a group $\op{T}\otimes
\op{G}$ with $\op{T}$ translational group of the axionic
symmetries this condition is certainly satisfied since
$m^{I\hat\L}$ is in the (dual) Lie algebra of $\op{T}$.\\
The expression for the potential looks quite complicated; however
we may reduce the given potential to known cases by suitable
erasing of some of the rows and columns of the matrices $\op{Q}$,
$\op{R}$, $\op{S}$.\\
First of all we note that if we delete the third row and the third
column of the given matrices we obtain the form of the potential
in the pure translational case with mass deformations given by
equation (\ref{eq:pot}) (with $\L \to \hat\L$). If we further
delete the first row and the first column, which means that we do
not implement mass deformations so that the tensor fields are
massless, then the potential reduces to \cite{Dall'Agata:2001zh}

\begin{equation}
\op{V} = -\mez \sx( \im \dx)^{-1|\hat \L \hat \S} e^I_{\hat \L}
e^J_{\hat \S} \o_I^x\o_J^x + 4 \sx( \op{M}_{IJ} - \o^x_I \o^x_J
\dx) e^I_{\hat\L} e^J_{\hat\S} \ol L^{\hat\L} L^{\hat\S}.
\end{equation}

\noi Another possibility is to delete the first two rows and the
first two columns, which means that we have no translational
gauging but only the gauging due to $\op{G}$. In this case the
potential takes the form

\begin{eqnarray}
\label{semisimple} \op{V}_{\op{G}} &=& g_{i\bar\jmath} k^i_{\check
\L} k^{\bar\jmath}_{\check \S} \ol L^{\check \L} L^{\check \S} + 4
\sx( g_{uv}+ A^I_u A^J_v \op{M}_{IJ} \dx) k^u_{\check \L}
k^v_{\check \S} \ol L^{\check \L} L^{\check \S} + \nn\\
&&- \sx(\mez \sx(\im\dx)^{-1|\check \L \check \S} + 4 \ol
L^{\check \L} L^{\check \S} \dx) {\cal P}^x_{\check \L} {\cal
P}^x_{\check \S}.
\end{eqnarray}

\noi Note that this potential has exactly the same form as in the
gauged standard $N=2$ supergravity coupled to vector multiplets
and hypermultiplets. The only difference is that the prepotentials
and the Killing vectors appearing in the last two terms are now
restricted to the manifold $\op{M}_T$ of the scalar tensor
multiplets. We also note that $g_{uv}+ A^I_u A^J_v \op{M}_{IJ}$ is
simply the original quaternionic metric restricted to the
$\s$--model. In an analogous way we could erase the first row and
the first column of the given matrices we obtain the potential
with gauging of group $\op{T} \otimes \op{G}$ but without mass
deformations and similarly for the other possibilities.

\section*{Acknowledgments}

We would like to thank Mario Trigiante for enlightening
discussions.\\
The work of R. D'Auria and L. Sommovigo has been supported by the
European Community's Human Potential Program under contract
HPRN-CT-2000-00131 Quantum Space--Time where R.D. and L.S. are
associated to Torino University.

\appendix
\section*{Appendix A: Completion of the Lagrangian and transformation laws}
\label{app:compl} \setcounter{equation}{0}\setcounter{section}{1}

In this Appendix we give the explicit expression of the 4--fermion
terms, $\op{L}_{\rm 4f}$ of the Lagrangian of section
\ref{Lagrangian} and the complete supersymmetry transformation
laws of the fermions including 3--fermion terms.\\
As observed in section \ref{Lagrangian}, the completion of the
4--fermion terms, $\op{L}_{\rm 4f}$ can be split in two pieces
namely $\op{L}_{\rm 4f}^{\rm non\, inv}$ and $\op{L}_{\rm 4f}^{\rm
inv}$.\\
$\op{L}_{\rm 4f}^{\rm non\, inv}$ can be fixed by the knowledge of
the Pauli terms, and as explained in the text, it is exactly the
same as $\op{L}_{\rm 4f}^{\rm non\, inv}$ of the standard $N=2$
supergravity \cite{Andrianopoli:1996ve}. We have:

\begin{eqnarray}
\op{L}_{\rm 4f}^{\rm non\,inv} & = & \sx\{ \sx( \im \dx)_{\L\S}
\sx[ 2 L^\L L^\S \sx( \ol \psi^A_\m \psi^B_\n \dx)^- \sx( \ol
\psi^C_\m \psi^D_\n \dx)^- \e_{AB} \e_{CD} + \dx.\dx. \nn\\
&& -8 {\rm i} L^\L \ol f^\S_{\bar\imath} \sx( \ol \psi^A_\m
\psi^B_\n \dx)^-
\sx( \ol \l^{\bar\imath}_A \g^\n \psi_B^\m \dx)^- + \nn\\
&& -2 \ol f^\L_{\bar\imath} \ol f^\S_{\bar\jmath} \sx( \ol
\l^{\bar\imath}_A \g^\n \psi_B^\m \dx)^- \sx( \ol
\l^{\bar\jmath}_C
\g_\n\psi_{D|\mu} \dx)^- \e^{AB}\e^{CD} + \nn\\
&& + \imez L^\L \ol f^\S_{\bar l} g^{k \bar l} C_{ijk} \sx( \ol
\psi^A_\m \psi^B_\n \dx)^- \ol \l^{iC} \g^{\m\n} \l^{jD} \e_{AB}
\e_{CD} + \nn \\
&& + \ol f^\L_{\bar m} \ol f^\S_{\bar l} g^{k \bar l} C_{ijk} \sx(
\ol \l^{\bar m}_A \g_\n \psi_{B|\m} \dx)^- \ol \l^{iA} \g^{\m\n}
\l^{jB} + \nn\\
&& - L^\L L^\S \sx( \ol \psi^A_\m \psi^B_\n \dx)^- \ol \z_\a
\g^{\m\n} \z_\b \, \e_{AB} \, \IC^{\a\b} + \nn\\
&& + {\rm i} L^\L \ol f^\S_{\bar\imath} \sx(\ol \l^{\bar\imath}_A
\g^\n \psi_B^\m \dx)^- \ol \z_\a \g_{\m\n} \z_\b \, \e^{AB} \,
\IC^{\a\b} + \nn \\
&& - \frac{1}{32} C_{ijk} \, C_{lmn} \, g^{k\bar r} \, g^{n\bar s}
\, \ol f^\L_{\bar r} \, \ol f^\S_{\bar s} \, \ol \l^{iA} \g_{\m\n}
\l^{jB} \, \ol \l^{kC} \g^{\m\n} \l^{lD} \e_{AB} \, \e_{CD} + \nn\\
&& -\frac{1}{8} L^\L \na_i f^\S_j \, \ol \z_\a \g_{\m\n} \z_\b \,
\ol \l^{iA} \g^{\m\n} \l^{jB} \e_{AB} \IC^{\a\b} + \nn \\
&& +\sx.\sx. \frac{1}{8} L^\L L^\S \, \ol \z_\a \g_{\m\n} \z_\b \,
\ol \z_\g \g^{\m\n} \z_\d \, \IC^{\a\b} \, \IC^{\g\d} \dx] + \cc
\dx\} \label{4fermnoninv}
\end{eqnarray}

\noi Of course the same terms can be obtained by supersymmetry.
The residual 4--fermion terms $\op{L}^{\rm non inv}_{\rm 4f}$ are
instead only fixed by supersymmetry. It is clear that this terms
are exactly the same as in the standard $N=2$ theory provided one
makes the following modifications:

\begin{enumerate}
\item First of all we have terms quadrilinear in the $\z_\a$
coming from the fact that dualization on space--time of the
axionic coordinates $q^I$ gives, together with the 3--form
$\op{H}_{I\m\n\r}$ also the bilinear $2 {\rm i} \, \D_I{}^\a{}_\b
\, \z_\a \g^\s \z^b \e_{\s\m\n\r}$.

\item One has also to reduce the symplectic curvature appearing in
the following 4--fermion term of $N=2$ standard supergravity

\begin{equation}
\mez \hat \op{R} (\hat \D)^\a{}_{\b \hat t \hat s}\, \op{U}^{\hat
t}{}_{A\g} \, \op{U}^{\hat s}_{B\d} \, \e^{AB} \, \IC^{\d\e} \,
\ol \z_\a \z_\e \, \ol \z^\b \z^\g
\end{equation}

\noi according to the rules explained in reference
\cite{Dall'Agata:2003yr} where $\hat \D^\a{}_\b$ is the Sp$(2n_H)$
connection on the quaternionic manifold $\op{M}_Q$

\begin{equation}
\hat \D_u^\a{}_\b = \D_u^\a{}_\b + A^I_u \D_I^\a{}_\b
\end{equation}

\end{enumerate}

\noi One obtains

\begin{eqnarray}
\hat \op{R}^\a{}_{\b t s} &=& \op{R}^\a{}_{\b t s} + F^I_{ts}
\D_I{}^\a{}_\b - A^I_{[t} \na_{s]} \D_I{}^\a{}_\b + A^I_t A^J_s \,
\D_I{}^\a{}_\g \D_J{}^\g{}_\b
\end{eqnarray}

\noi where the hatted indices in the l.h.s. refer to the
quaternionic manifold, $F^I_{ts} = \der_{[t} A^I_{s]}$ and
$\op{R}^\a{}_{\b t s}$ is the symplectic curvature on $\op{M}_T$
defined by

\begin{equation}
\op{R}^\a{}_{\b} = {\rm d} \D^\a{}_\b + \D^\a{}_\g \wedge
\D^\g{}_\b.
\end{equation}

\noi Taking into account the previous modifications $\op{L}_{\rm
4f}^{\rm inv}$ becomes

\begin{eqnarray}
\op{L}_{\rm 4f}^{\rm inv} & = & \imez \sx( g_{i\bar\jmath} \, \ol
\l^{iA} \g_\s \l^{\bar\jmath}_B - 2 \d^A_B \, \ol \z^\a \g_\s
\z_\a \dx) \ol \psi_{A|\m} \g_\l \psi^B_\n \,
\frac{\e^{\m\n\l\s}}{\sqrt{-g}} + \nn \\
&& - \frac{1}{6} \sx( C_{ijk} \, \ol \l^{iA} \g^\m \psi^B_\m \,
\ol \l^{jC} \l^{kD} \e_{AC} \e_{BD} +\cc \dx) + \nn \\
&& - 2 \ol \psi^A_\m \psi^B_\n \ol \psi_A^\m \psi_B^\n + 2
g_{i\bar\jmath} \, \ol \l^{iA} \g_\m \psi^B_\n \, \ol
\l^{\bar\imath}_A \g^\m \psi_B^\n + \nn\\
&& +\frac{1}{4} \sx( R_{i\bar\jmath l\bar k} + g_{i\bar k} \,
g_{l\bar\jmath} - \frac{3}{2} g_{i\bar\jmath} \, g_{l\bar k} \dx)
\ol \l^{iA} \l^{lB} \, \ol \l^{\bar\jmath}_A \l^{\bar k}_B \nn + \\
&& + \frac{1}{4} g_{i\bar\jmath} \, \ol \z^\a \g_\m \z_\a \ol
\l^{iA} \g^\m \l^{\bar\jmath}_A + \mez \sx( \op{R}^\a{}_{\b t s} +
F^I_{ts} \D_I{}^\a{}_\b \dx.+ \nn\\
&& \sx. - A^I_{[t} \na_{t]} \D_I{}^\a{}_\b + A^I_t A^J_s
\D_I{}^\a{}_\g \D_J{}^\g{}_\b\dx) \e^{AB} \, \IC^{\d\eta} \, \ol
\z_\a \z_\eta \, \ol \z^\b \z^\g + \nn \\
&& - \sx[ \frac{{\rm i}}{12} \na_m C_{jkl} \ol \l^{jA} \l^{mB} \ol
\l^{kC} \l^{lD} \e_{AC} \e_{BD} + \cc \dx] + \nn \\
&& + g_{i\bar\jmath} \, \ol \psi^A_\m \l^{\bar\jmath}_A \, \ol
\psi_B^\m \l^{iB} + 2 \ol \psi^A_\m \z^\a \, \ol \psi_A^\m \z_\a
+\nn\\
&& + \sx( \e_{AB} \, \IC_{\a\b} \, \ol \psi^A_\m \z^\a \,\ol
\psi^{B|\m} \z^\b + \cc \dx) + \nn\\
&& + 120 \op{M}^{IJ} \D_I{}^\a{}_\b \, \D_J{}^\g{}_\d \, \z_a
\g^\m \z^\b \, \z_\g \g_\m \z^\d + \nn\\
&& + 24\, {\rm i}\, \op{M}^{IJ} \D_I{}^\a{}_\b \, \z_\a \g_\m
\z^\b \sx[ \op{U}_J{}^{A\g} \ol \psi_{A|\n} \g^{\m\n} \z_\g +
\op{U}_{J\,A\g} \ol \psi^A_\n \g^{\m\n} \z^\g \dx] + \nn\\
&& + 24\, {\rm i}\, \op{M}^{IJ} \D_I{}^\a{}_\b\, \z_\a\g^\m \z^\b
\sx[ \op{U}_J{}^{A\g} \ol \psi_{A|\m} \z_\g + \op{U}_{J\,A\g} \ol
\psi^A_\m \z^\g \dx] \label{4ferminv}.
\end{eqnarray}

\noi The supersymmetry transformation laws of the fermions turn
out to be

\begin{eqnarray}
\d\psi_{A|\m} &=& \op{D}_\m \e_A - \frac{1}{4} \sx( \der_i K \,
\ol \l^{iB} \ve_B - \der_{\bar\imath} K \, \ol \l^{\bar\imath}_B
\ve^B \dx) \psi_{A\m} + \nn\\
&& -h^I_\m \o_{IA}{}^B \ve_B + 12 \, {\rm i} \, \D^{I\a}{}_\b
\z_\a \g_\m \z^\b \o_{IA}{}^B \ve_B + \nn\\
&&-2 \o_{u\,A}{}^B P^u{}_{C\a} \sx( \e^{CD}\, \IC^{\a\b} \, \ol
\z_\b \ve_D + \ol \z^\a \e^C \dx) \psi_{B\m} + \nn\\
&&-2 \o_{I\,A}{}^B \op{U}^I{}_{C\a} \sx( \e^{CD}\, \IC^{\a\b}\,
\ol \z_\b \ve_D + \ol \z^\a \ve^C \dx)\psi_{B\m} + \nn\\
&& + \sx( A_A{}^{\n B} \eta_{\m\n} + A'_A{}^{\n B} \g_{\m\n}
\dx) \ve_B + \nn\\
&& + \sx[ {\rm i} S_{AB} \eta_{\m\n} + \e_{AB} \sx(
T^-_{\m\n} + U^+_{\m\n} \dx) \dx] \g^\n \ve^B \label{trasfgrav}\\
\d\l^{iA} &=& \frac{1}{4} \sx( \der_j K \, \ol \l^{jB} \e_B -
\der_{\bar\jmath} K \, \ol \l^{\bar\jmath}_B \ve^B \dx)\l^{iA}+ \nn\\
&&-2 \o_I{}^A{}_B \op{U}^I{}_{C\a} \sx( \e^{CD}\, \IC^{\a\b}\, \ol
\z_\b \e_D + \ol \z^\a \e^C \dx) \l^{iB} + \nn\\
&& - \o_u{}^A{}_B P^u{}_{C\a} \sx( \e^{CD} \, \IC^{\a\b} \, \ol
\z_\b \e_D + \ol \z^\a \ve^C \dx) \l^{iB} + \nn\\
&&- \G^i{}_{jk}\, \ol \l^{kB} \ve_B\, \l^{jA} + {\rm i}\sx(\na_\mu
z^i - \ol \l^{iA} \psi_{A|\m} \dx) \g^\m \ve^A + \nn\\
&&+G^{-i}_{\m\n}\, \g^{\m\n} \ve_B \e^{AB} + D^{iAB} \ve_B
\label{gaugintrasfm}\\
\d\z_\a &=& - \D_{u\,\a}{}^\b P^u{}_{\g A} \sx( \e^{AB}\,
\IC^{\g\d}\, \ol \z_\d \ve_B + \ol \z^\g \ve^A \dx)\z_\b + \nn\\
&& -h^I_\m \o_{IA}{}^B \ve_B - 12\,{\rm i} \, \D^{I\a}{}_\b \z_\a
\g_\m \z^\b \o_{IA}{}^B \ve_B + \nn\\
&&+ \frac{1}{4} \sx( \der_i K \, \ol \l^{iB} \ve_B -
\der_{\bar\imath} K \, \ol \l^{\bar\imath}_B \ve^B \dx) \z_\a + \nn\\
&&+ {\rm i} \sx( P_u{}^{B\b} \na_\m q^u - \e^{BC}\, \IC^{\b\g}\,
\ol \z_\g \psi_{C|\m} - \ol \z^\b \psi^B_\m \dx) \g^\m \ve^A
\e_{AB} \, \IC_{\a\b} + \nn\\
&& - {\rm i} h^I_\m \op{U}_I{}^{B\b} \g^\m \ve^A \e_{AB}\,
\IC_{\a\b} + N_\a^A \ve_A. \label{iperintrasf}
\end{eqnarray}

\noi Note that these transformation laws, which have been obtained
from the solution of Bianchi identities in superspace (see
Appendix B), could be obtained from the corresponding formulae of
$N=2$ standard supergravity of reference
\cite{Andrianopoli:1996ve} by reducing the quaternionic index as
explained in reference \cite{Dall'Agata:2003yr} provided we make a
suitable Ansatz in superspace for the 1--form $h_I$ as explained
in the next Appendix.

\section*{Appendix B: Solution of Bianchi identities and the construction of the rheonomic Lagrangian.}
\setcounter{equation}{0} \label{app:biid}

In this Appendix we first describe the geometric approach for the
derivation of the $N=2$ supersymmetry transformation laws of the
physical fields, and then we construct the rheonomic superspace
Lagrangian. The solution of Bianchi identities in superspace will
provide us with the supersymmetry transformation laws in
space--time, while the restriction of the rheonomic Lagrangian to
space--time will give the Lagrangian of section \ref{Lagrangian}
and its completion given in Appendix A. Since the present approach
is completely analogous to that used for $N=2$ standard
supergravity we will be very short referring for more
details to Appendices A and B of the quoted reference.\\
The first step to perform is to extend the physical fields to
superfields in $N=2$ superspace: that means that the space--time
2--form $B_I$, 1--forms $\o^{a\,b}$, $V^a$,$\psi^A$, $\psi_A$,
$A^\L$, and the space--time 0--forms $\l^{iA}$,
$\l_A^{\bar\imath}$, $z^i$, $\ol z^{\bar\imath}$, $\z_\a$,
$\z^\a$, $q^u$ are promoted to 2--, 1-- and 0--superforms in
$N=2$ superspace, respectively.\\
The definition of the superspace "curvatures" in the gravitational
and vector multiplet sectors is identical to that of standard
$N=2$ supergravity except for the fact that the composite 1--form
connections $\op Q$, $\o^A{}_B$ and $\D^\a{}_\b$ are now
restricted to $\op{M}_T$ instead of the quaternionic manifold.\\
We have:

\begin{eqnarray}
T^a &\equiv& {\rm d} V^a - \o^a{}_b \wedge V^b - {\rm i}\, \ol
\psi_A \wedge \g^a \psi^A = 0\label{torsdef}\\
\r_A &\equiv& {\rm d} \psi_A - \frac{1}{4} \g_{ab} \, \o^{ab}
\wedge \psi_A + \imez \op{Q} \wedge \psi_A + \o_A{}^B \wedge
\psi_B
\equiv \na \psi_A \label{gravdefdown} \\
\r^A &\equiv& {\rm d} \psi^A - \frac{1}{4} \g_{ab} \, \o^{ab}
\wedge \psi^A - \imez \op{Q} \wedge \psi^A + \o^A{}_B \wedge
\psi^B
\equiv \na \psi^A \label{gravdefup} \\
R^{ab} &\equiv& {\rm d} \o^{ab} - \o^a{}_c \wedge \o^{cb}
\label{riecurv}\\
\na z^i &=& {\rm d} z^i \label{zcurv}\\
\na \bar z^{\bar\imath} &=& {\rm d}{\bar z}^{\bar\imath} \label{zcurvb}\\
\na \l^{iA} &\equiv& {\rm d} \l^{iA} - \frac{1}{4} \g_{ab} \,
\o^{ab} \l^{iA} - \imez \op{Q} \l^{iA} + \G^i{}_j \l^{jA} +
\o^A{}_B \l^{iB} \label{lamcurv} \\
\na\l^{\bar\imath}_A &\equiv& {\rm d} \l^{\bar\imath}_A -
\frac{1}{4} \g_{ab} \,\o^{ab} \l^{\bar\imath}_A + \imez \op{Q}
\l^{\bar\imath}_A + \G^{\bar\imath}{}_{\bar\jmath}
\l^{\bar\jmath}_A + \o_A{}^B
\l^{\bar\imath}_B \label{lamcurvb}\\
F^\L &\equiv& {\rm d} A^\L + \ol L^\L \ol \psi_A \wedge \psi_B
\e^{AB} + L^\L \ol \psi^A \wedge \psi^B \e_{AB} \label{Fcurv}
\end{eqnarray}

\noi where $\o_A{}^B = \imez \o^x \s^x{}_A{}^B$ ($\o^A{}_B =
\e^{AC} \e_{DB}\, \o_C{}^D$) and $\op{Q}$ are connections on the
SU$(2)$ bundle defined on $\op{M}_T$ and on the U$(1)$ bundle
defined on $\op{M}_{SK}$, namely $\op{Q} = -\imez (\der_i K {\rm
d}z^i - \der_{\bar\imath} K {\rm d}z^{\bar\imath})$ with $K$ being
the K\"ahler potential. The index $\L$ runs from $0$ to $n_V$, the
index $0$ referring to the graviphoton. The Levi--Civita
connection on $\op{M}_{SK}$ is $\G^i{}_j$ and $L^\L(z, \ol z) =
e^{\frac{\op K}{2}} X^\L(z)$ is the upper half (electric) of the
symplectic section of special geometry. Note that equation
(\ref{torsdef}) is a superspace
constraint.\\
Instead of the hypermultiplet curvatures we have now the tensor
multiplet curvatures defined as

\begin{eqnarray}
H_I&\equiv& {\rm d}B_I - \o_{IA}{}^B \, \ol \psi^A \wedge \g_a
\psi_B \wedge V^a \label{bcurv}\\
P^{A\a} &\equiv& P_u{}^{A\a} {\rm d} q^u\label{ucurv}\\
\na \z_\a &\equiv& {\rm d} \z_\a - \frac{1}{4} \o^{ab} \g_{ab}
\z_\a - \imez \op{Q} \z_\a + \D_\a{}^\b \z_\b \label{iperincurv}\\
\na \z^\a &\equiv& {\rm d} \z^\a - \frac{1}{4} \o^{ab} \g_{ab}
\z^\a + \imez \op{Q} \z^\a + \D^\a{}_\b \z^\b. \label{iperincurvb}
\end{eqnarray}

\noi Applying the d operator to equations
(\ref{torsdef})--(\ref{iperincurvb}), one finds:

\begin{eqnarray}
&\op{D} T^a & + R^{ab} \wedge V^b - {\rm i} \ol \psi^A \wedge
\g^a \r_A + {\rm i} \ol \r^A \wedge \g^a \psi_A = 0 \label{torsbi}\\
&\na \r_A& + \frac{1}{4} \g_{ab} R^{ab} \wedge \psi_A - \imez K
\wedge \psi_A - \imez R_A{}^B \wedge \psi_B = 0 \label{gravbidown} \\
&\na \r^A& + \frac{1}{4} \g_{ab} R^{ab} \wedge \psi^A + \imez K
\wedge \psi^A + \imez R^A{}_B \wedge \psi^B = 0 \label{gravbiup} \\
&\op{D} R^{ab}& = 0 \label{riebi}\\
&\na^2 z^i& = 0 \label{zbi}\\
&\na^2 z^{\bar\imath}& = 0 \label{zstarbi}\\
&\na^2 \l^{iA}& + \frac{1}{4} \g_{ab} R^{ab} \l^{iA} + \imez K
\l^{iA}+ R^i{}_j \l^{jA} - \imez R^A{}_B \l^{iB} = 0 \label{lambi}\\
&\na^2 \l^{\bar\imath}_A& + \frac{1}{4} \g_{ab} R^{ab}
\l^{\bar\imath}_A - \imez K \l^{\bar\imath}_A +
R^{\bar\imath}{}_{\bar\jmath} \l^{\bar\jmath}_A - \imez R_A{}^B
\l^{\bar\imath}_B = 0 \label{lamstarbi}\\
&\na \hat F^\L& - \na \ol L^\L \wedge \ol \psi_A \wedge \psi_B
\e^{AB} - \na L^\L \wedge \ol \psi^A \wedge \psi^B \e_{AB} \nn\\
&& + 2 \ol L^\L \ol \psi_A \wedge \r_B \e^{AB} + 2  L^\L \ol
\psi^A \wedge \r^B \e_{AB} + \nn\\
&& + 2 m^{I\L} \sx( H_I + \o_{I\,A}{}^B \ol \psi^A \g_a \psi_B
\wedge V^a \dx)= 0 \label{Fbi}\\
&\na H_I& - \na \o_{I\,A}{}^B \wedge \ol \psi_B \wedge \g_a \psi^A
\wedge V^a + \o_{I\,A}{}^B \sx( \ol \psi_B \wedge \g_a \r^A + \dx.
\nn \\
&&\sx. + \ol\psi^A \wedge \g_a \r_B \dx)\wedge V^a - {\rm i}\,
\o_{I\,A}{}^B \ol \psi_B \wedge \g_a \psi^A \wedge
\ol \psi_C \wedge \g^a \psi^C \label{Hbi}\\
&\na P^{A\a}& = 0 \label{Ubi}\\
&\na^2 \z_\a& + \frac{1}{4} R^{ab} \g_{ab} \z_\a + \imez K \z_\a
+ R_\a{}^\b \z_\b = 0 \label{iperibi}\\
&\na^2 \z^\a& + \frac{1}{4} R^{ab} \g_{ab} \z^\a - \imez K \z^\a +
R^\a{}_\b \z^\b = 0 \label{iperibib}
\end{eqnarray}

\noi where the covariant derivatives are defined by equations
(\ref{gravdefdown}), (\ref{gravdefup}), (\ref{lamcurv}),
(\ref{lamcurvb}), (\ref{iperincurv}) and (\ref{iperincurvb}).\\
The solution can be obtained as follows: first of all one requires
that the expansion of the curvatures along the intrinsic
$p$--forms basis in superspace, namely: $V$, $\psi$, $V\wedge V$,
$V\wedge\psi$, $\psi \wedge \psi$, $V \wedge V \wedge V$,
$V^a\wedge V^b\wedge \psi$, $V^a \wedge \psi \wedge \psi$ and
$\psi \wedge \psi \wedge \psi$, is given in terms only of the
physical fields (rheonomy). This
insures that no new degree of freedom is introduced in the theory.\\
Secondly one writes down such expansion in a form which is
compatible with all the symmetries of the theory, that is:
covariance under U$(1)$--K\"ahler and
SU$(2)\otimes$Sp$(2,m)\otimes$SO$(n_T)$, Lorentz transformations
and reparametrization of the scalar manifold $\op{M}_{SK}\otimes
\op{M}_T$. This fixes completely the Ansatz for the curvatures at
least if we exclude higher derivative interactions (for a more
detailed explanation the interested reader is referred to the
standard reference of the geometrical approach \cite{CaDFb} and
to the Appendices A and B of reference \cite{Andrianopoli:1996ve}).\\
It is important to note that, in order to satisfy the gravitino
Bianchi identity in the $\psi \wedge \psi \wedge \psi$ sector, the
second term in equation (\ref{paramgrav}), even though it is a
3--fermion term, is essential since its presence allows the
cancellation of the shift term $S_{AB}$ in the gravitino
parametrization (in the usual gauged $N=2$ standard supergravity
the shift $S_{AB}$ is instead cancelled by the additional terms in
the gauged curvature $R^{ab}$).\\
The final parametrizations of the superspace curvatures, are given
by:

\begin{eqnarray}
T^a &=& 0 \label{paramtors}\\
\r_A &=& \tilde \r_{A| ab} V^a \wedge V^b - \o_{IA}{}^B
\op{U}^I{}_{C\a} \sx( \e^{CD} \IC^{\a\b}\, \ol \z_\b \psi_D
+ \ol \z^\a \psi^C\dx)\wedge \psi_B + \nn\\
&& - h^I_a \,\o_{IA}{}^B \psi_B \wedge V^a + \sx ( A_A{}^{B|b}
\eta_{ab} + A'_A{}^{B|b}\g_{ab} \dx) \psi_B \wedge V^a\nn\\
&& + \sx[ {\rm i} S_{AB}\eta_{ab}+ \e_{AB}( T^-_{ab}+ U^+_{ab})
\dx] \g^b \psi^B \wedge V^a \label{paramgrav}\\
R^{ab} &=& \tilde{R}^{ab}{}_{cd} V^c\wedge V^d - {\rm i} \sx( \ol
\psi_A \th^{A|ab}_c + \ol \psi^A \th^{ab}_{A|c} \dx) \wedge V^c + \nn\\
&& + \e^{abcf} \ol\psi^A \wedge \g_f \psi_B \sx(A'^B{}_{A|c}-\ol
A'_{A|c}{}^B \dx) + \nn\\
&& - 2 \,{\rm i} \, \o_{I\,A}{}^B \ol \psi^A \g^a \psi_B\, \sx(
h^{I\,b} +12\, {\rm i} \, \D_I{}^\a{}_\b \, \ol \z_\a \g^b \z^\b\dx)+ \nn\\
&& + {\rm i} \e^{AB} \sx( T^{+ab} + U^{-ab} \dx) \ol \psi_A \wedge
\psi_B - {\rm i} \e_{AB}  \sx( T^{-ab} + U^{+ab} \dx) \psi^A \wedge \psi^B\nn\\
&& - S_{AB} \ol \psi^A \wedge \g^{ab} \psi^B - \ol S^{AB} \ol
\psi_A \wedge \g^{ab} \psi_B\label{paramriem}\\
\hat F^\L &=& \tilde{\hat{F}}^\L_{ab} V^a \wedge V^b+ {\rm i} \sx(
f^\L_i \ol \l^{iA} \g_a \psi^B \e_{AB} + {\rm i} \ol
f^\L_{\bar\imath} \ol\l^{\bar\imath}_A \g_a \psi_B \e^{AB} \dx)
\wedge V^a \label{gaugparam}\\
\na \l^{iA} &=& \tilde{\na_a \l}^{iA} V^a+ {\rm i}\, \tilde Z^i_a
\g^a \psi^A + G^{-i}_{ab} \g^{ab} \e^{AB} \psi_B + W^{iAB} \psi_B \label{gauginparam}\\
\na z^i &=& \tilde Z^i_a V^a + \ol \l^{iA} \psi_A \label{scalparam} \\
P^{A\a} &=& \tilde P_a{}^{A\a} V^a +\ol\psi^A\z^\a +\e^{AB}
\IC^{\a\b}\, \ol \psi_B\z_\b\label{pparam}\\
H_I &=& \tilde H_{Iabc} V^a \wedge V^b \wedge V^c -\imez \sx(
\op{U}_I{}^{A\a} \ol \psi_A \g_{ab} \z_\a - \op{U}_{IA\a} \ol
\psi^A \g_{ab} \z^\a
\dx) \wedge V^a \wedge V^b \label{hparam}\\
\na\z_\a &=& \tilde{\na_a \z}_\a V^a + {\rm i} P_{a\,A\a} \g^a
\psi^A - {\rm i} h^I_a \, \op{U}_{I\,A\a}\, \g^a \psi^A + N_\a^A
\psi_A \label{iperinparam}
\end{eqnarray}

\noi where:

\begin{eqnarray}
A_A{}^{|a B} &=& - \frac{{\rm i}}{4} \, g_{\bar k l}\, \sx( \ol
\l^{\bar k}_A \g^a \l^{l B} - \d^B_A \ol \l^{\bar k}_C \g^a \l^{l C}
\dx)\label{Adefrheo}\\
A'_A{}^{|a B} &=& \frac{{\rm i}}{4} \, g_{\bar k l}\, \sx( \ol
\l^{\bar k}_A \g^a \l^{l B} - \mez \d^B_A \ol \l^{\bar k}_C \g^a
\l^{l C} \dx) - \frac{{\rm i}}{4} \d_A^B \ol \z_\a \g^a
\z^\a\label{A'defrheo}\\
\th^{ab|c}_A &=& 2 \g^{[a} \r^{b]c}_A + \g^c \r^{ab}_A; \quad
\th^{ab\,A}_c = 2 \g^{[a} \r^{b]c|A} + \g^c \r^{ab|A} \label{thetadef}\\
T^-_{ab} &=& \sx(\op{N} - \ol \op{N} \dx)_{\L\S} L^\S \sx( \tilde{
\hat{F}}_{ab}^{\L-} + \frac{1}{8} \na_i f^\L_j \ol \l^{i A}
\g_{ab} \l^{jB} \e_{AB} \dx.\nn\\
&& \qquad\qquad\sx. - \frac{1}{4} \IC^{\a\b} \ol \z_\a \g_{ab}
\z_\b L^\L \dx) \label{T-defrheo}\\
U^-_{ab} &=& \frac{{\rm i}}{4} \l \IC^{\a\b} \ol \z_\a \g_{ab}
\z_\b \label{U-defrheo}\\
G^{i-}_{ab} &=& \frac{{\rm i}}{2} g^{i\bar \jmath} \ol f^\G_{\bar
\jmath} \sx( \op{N} -\ol \op{N} \dx)_{\G\L} \sx( \tilde{\hat{
F}}^{\L-}_{ab} + \frac{1}{8} \na_k f^\L_l \ol \l^{kA} \g_{ab}
\l^{l B} \e_{AB} \dx.\nn\\
&& \qquad\qquad\sx. - \frac{1}{4} \IC^{\a\b} \ol\z_\a \g_{ab}
\z_\b L^\L \dx) \label{G-defrheo}\\
S_{AB} &=& \imez \s^x{}_{AB} \o^x_I \sx( e^I_\L L^\L - m^{I\L}
M_\L \dx) \label{Sdefrheo}\\
N_\a^A &=& - 2 \, \op{U}_I{}^A{}_\a \sx( e^I_\L L^\L - m^{I\L}
M_\L \dx) \label{Ndefrheo}\\
W^{iAB}&=& {\rm i}\, g^{i \bar\jmath}\, \s^{x\,AB} g \op{P}^x_\L
\ol f_{\bar\jmath}^\L + \nn\\
&&\quad + {\rm i}\, g^{i \bar\jmath}\, \s^{x\,AB} \o^x_I
\sx(e^I_\L \ol f^\L_{\bar\jmath} - m^{I\L} \ol h_{\L\bar\jmath}
\dx)\label{Wdefrheo}
\end{eqnarray}

\noi where according to what has been discussed in the main text,
$\hat F^\L = F^\L + 2 m^{I\L} B_I$.\\
Let us now discuss the appearance in the parametrization of the
fermions of the shift terms $S_{AB}$, $W^{iAB}$ and $N^A_\a$.\\
If we set these shifts to zero, we have the solution of the
Bianchi identities provided $\hat\op{F}$ is replaced by $\op{F}$.
The extended solution containing these shifts can be retrieved as
follows: if we look at the Bianchi identity (\ref{Fbi}), with
$\hat\op{F}$ replaced by $\op{F}$, then adding the so far unknown
shifts gives in the sector $\ol \psi^B \g^a \psi_C V^b$ the
following extra terms:

\begin{equation}\label{extra}
\sx(-\mez \e_{AB} W^{iAC} f^\L_i + L^\L S^{AC} \e_{BA} + \cc \dx)
2\,{\rm i}\, \ol \psi^B \wedge \g_a \psi_C \wedge V^a = 0.
\end{equation}

\noi Taking into account the identities of Special Geometry
\cite{Andrianopoli:1996ve}, the solution of equation (\ref{extra})
must have the following form:

\begin{eqnarray}
S^{AC} &=& D^{AC}_\L L^\L\label{sac}\\
W^{iAC} &=& 2 g^{i\bar\jmath} f^\L_{\bar \jmath} D^{AC}_\L
\label{wiac}
\end{eqnarray}

\noi where $D^{AC}_\L$ is a set of $n_V+1$ SU$(2)$ symmetric
tensors so far undetermined. The magnetic mass--shifts can be
obtained as follows; we set:

\begin{equation}
D^{AC}_\L = -\imez \s^{x\,AC} \o^x_I m^{I\L} M_\L \label{dac}
\end{equation}

\noi then using the Special Geometry identity:

\begin{equation}
g^{\bar\imath j} f^\L_{\bar \imath} h_{\S j} - \cc = {\rm i}
\d^\L_\S + \ol L^\L M_\S - L^\L \ol M_\S \label{spgeo}
\end{equation}

\noi one sees that the residual term

\begin{equation}
- m^{I\L} M_\L \o_I{}_A{}^B \ol \psi^A \wedge \g_a \psi_B \wedge
V^a \label{resterm}
\end{equation}

\noi is cancelled by the redefinition (\ref{redeff1}) using
equation (\ref{bcurv}). This solution for the magnetic shifts can
be further generalized to include electric mass--shifts by
extending the definition of $D^{AC}_\L$ as follows:

\begin{equation}
D^{AC}_\L \to D^{AC}_\L + \imez \s^{x\,AC} \o^x_I e^I_\L L^\L.
\end{equation}

\noi Indeed the new terms cancel identically by virtue of the
Special Geometry formula

\begin{equation}
g^{\bar\imath j} f_{\bar\imath}^\L f^\S_j = -\mez
\sx(\im\dx)^{-1|\L\S} - L^\L \ol L^\S.
\end{equation}

\noi Finally the mass--shift $N_\a^A$ can be computed from the
gravitino Bianchi identity looking at the coefficients of the
$\psi\wedge\psi\wedge\psi$ terms. Indeed in this case one finds,
after a Fierz rearrangement, the following equation:

\begin{equation}
S_{AB} = -\frac{\rm i}{4} \s^x{}_{AB} \o^x_I \op{U}^{IC}{}_\a
N^\a_C
\end{equation}

\noi which determines $N^\a_A$ to be

\begin{equation}
N^\a_A = -2\, \op{U}_{IA}{}^\a \sx( e^I_\L L^\L - m^{I\L} M_\L
\dx).
\end{equation}

\noi where we have used the identity \cite{Dall'Agata:2003yr}:

\begin{equation}
\op{U}_{IA}{}^\a \op{U}_{JB\a} + \op{U}_{JA}{}^\a \op{U}_{IB\a} =
\op{M}_{IJ} \e_{AB}.
\end{equation}

\noi In this way equations (\ref{Sdefrheo}), (\ref{Wdefrheo}) and
(\ref{Ndefrheo}) are reproduced.\\
It is important to stress that the field--strengths
$\tilde{R}^{ab}{}_{cd}$, $\tilde{\r}_{A|ab}$,
$\tilde{F}^\L_{ab}$,\break $\tilde{P}^{A\a}_a \equiv P^{A \a}_u
\tilde{\na_a q}^u$, $\tilde{\na_a\l}^{iA}$, $\tilde{\na_a\z}_\a$
and their hermitian conjugates are not space--time
field--strengths since they are components along the bosonic
vielbeins $V^a=V^a_\m\,dx^\m + V^a_\a {\rm d} \th^\a$ where
($V^a_\m$, $V^a_\a$) is a submatrix of the super--vielbein matrix
$E^I \equiv$($V^a$, $\psi$) (see Appendix A of reference
\cite{Andrianopoli:1996ve}). Note that in the component approach
the "tilded" field--strengths defined in the previous equations
are usually referred to as the "supercovariant"
field--strengths.\\
The previous formulae refer to the theory without gauging of the
group $\op{G}$. If the gauging is turned on then we must use
gauged quantities according to the rules explained in reference
\cite{Andrianopoli:1996ve}. In particular the differentials on the
scalar fields $z^i$, $\ol z^{\bar \imath}$ and $q^u$ are redefined
as in equation (\ref{gaugzder}), (\ref{gaugzbarder}),
(\ref{gaugzbarder}) and the composite connections and curvatures
acquire extra terms as explained in reference
\cite{Andrianopoli:1996ve}. Furthermore for the gauged theory the
index $\L$ has to be split according to the discussion of
subsection \ref{adding}. For example, equation (\ref{Ubi}),
(\ref{zbi}) and (\ref{zstarbi}) become:

\begin{eqnarray}
&\na P^{A\a}& - \sx( F^{\check\L} - \ol L^{\check\L} \ol \psi_A
\psi_B \e^{AB}- L^{\check\L} \ol \psi^A \psi^B \e_{AB}\dx)
k^u_{\check\L} P_u{}^{A\a} = 0\label{qbinew}\\
&\na^2 z^i& - \sx( F^{\check\L} - \ol L^{\check\L} \ol \psi_A
\wedge \psi_B \e^{AB} - L^{\check\L} \ol \psi^A \wedge \psi^B
\e_{AB} \dx) k_{\check\L}^i = 0 \label{zbinew}\\
&\na^2 z^{\bar\imath}&- \sx( F^{\check\L} - \ol L^{\check\L} \ol
\psi_A \wedge \psi_B \e^{AB} - L^{\check\L} \ol \psi^A \wedge
\psi^B \e_{AB} \dx) k_{\check\L}^{\bar\imath}= 0.
\label{zstarbinew}
\end{eqnarray}

\noi Furthermore the curvature 2--forms $K$, $R^{ab}$, $R^{\a\b}$
become gauged according to the procedure of reference
\cite{Andrianopoli:1996ve}.\\
Finally we recall that the solution we found is an on--shell
solution, and it also determines the geometry of the Special
K\"ahler manifold (see Appendix A of \cite{Andrianopoli:1996ve})
and the geometry of $\op{M}_T$ that has been analyzed in reference
\cite{Dall'Agata:2003yr}.\\
The determination of the superspace curvatures enables us to write
down the $N=2$ supersymmetry transformation laws. Indeed we recall
that from the superspace point of view a supersymmetry
transformation is a Lie derivative along the tangent vector:

\begin{equation}
\e = \ol \e^A \vec D_A + \ol \e_A \vec D^A
\end{equation}

\noi where the basis tangent vectors $\vec D_A$, $\vec D^A$ are
dual to the gravitino 1--forms:

\begin{equation}
\vec D_A \sx(\psi^B \dx) = \vec D^A \sx(\psi_B \dx) =\bfone
\end{equation}

\noi where $\bfone$ is the unit in spinor space.\\
Denoting by $\m^{\op I}$ and $R^{\op I}$ the set of 1-- and
2--forms $\sx( V^a\dx.$, $\psi_A$, $\psi^A$, $A^\L$, $\sx. B_I
\dx)$ and of the 2-- and 3--forms $\sx( R^{ab} \dx.$, $\r_A$,
$\r^A$, $F^\L$, $\sx. H_I \dx)$ respectively, one has:

\begin{equation}
\ell \mu^{\op I} = \sx(i_{\e} {\rm d}\,+\,{\rm
d}i_{\e}\dx)\mu^{\op I} \equiv \sx(D \e \dx)^{\op I} \,+
\,i_{\e}R^{\op I}
\end{equation}

\noi where $D$ is the derivative covariant with respect to the
$N=2$ Poincar\'e superalgebra and $i_\e$ is the contraction
operator along the tangent vector $\e$.\\
In our case:

\begin{eqnarray}
\sx(D \e \dx)^a &=& {\rm i} \sx( \ol \psi_A \g^a \e^A + \ol \psi^A
\g^a \e_A \dx)\\
\sx(D \e\dx)^\a &=& \na \e^\a\\
\sx(D \e\dx)^\L &=& 2 \ol L^\L \ol \psi_A \e_B \e^{AB} + 2 L^\L \ol
\psi^A \e^B \e_{AB}\\
\sx(D \e \dx)_{\op I}&=& \o_{I\, A}{}^B \sx( \op \psi_B \g_a \e^A
+ \ol \psi^A \g_a \e_B \dx) V^a
\end{eqnarray}

\noi (here $\a$ is a spinor index)\\
For the 0--forms which we denote shortly as $\n^{\op I} \equiv
\sx( q^u\dx.$, $z^i$, $\ol z^{\bar\imath}$, $\l^{iA}$,
$\l^{\bar\imath}_A$, $\z_\a$, $\z^\a$ we have the simpler result:

\begin{equation}
\ell_{\e} = i_\e {\rm d} \nu^{\op I} = i_\e \sx( \na \n^{\op I} -
{\rm connection\,terms}. \dx)
\end{equation}

\noi Using the parametrizations given for $R^{\op I}$ and
$\na\n^{\op I}$ and identifying $\d_\e$ with the restriction of
$\ell_\e$ to space--time it is immediate to find the $N=2$
supersymmetry laws for all the fields. The explicit formulae are
given in Appendix
A.\\
~\\
\noi Let us now derive the space--time Lagrangian from the
rheonomic Lagrangian. The meaning of the rheonomic Lagrangian has
been discussed in detail in the literature. In our case one can
repeat almost verbatim, apart of some minor modifications, the
considerations given in Appendix B of \cite{Andrianopoli:1996ve}.
We limit ourselves to write down the rheonomic Lagrangian up to
4--fermion terms since the 4--fermion terms are almost immediately
reconstructed from their space--time expression given in Appendix
A by promoting the various terms to 4--forms in superspace. The
most general Lagrangian (up to 4--fermion terms) has the following
form (wedge products are omitted):

\begin{equation}
\op{A} = \int_{\op{M}^4 \subset \op{M}^{4|8}} \op{L}.
\end{equation}

\begin{eqnarray}
  \label{eq:rhlaggrav}
  \op{L}_{\rm Grav}&=& R^{ab} V^c  V^d \e_{abcd}- 4 \sx(\ol
  \psi^A  \g_a \r_A - \ol\psi_A  \g_a \r^A\dx)  V^a\\
  \label{eq:rhlagkin}
  \op{L}_{\rm kin} &=&\b_1 g_{i\bar\jmath}\sx[{\bf Z}^i_a\sx(\na \ol z^{\bar\jmath}
  -\ol\psi^A \l^{\bar \jmath}_A\dx) +{\bf \ol Z}^{\bar \jmath}_a \sx( \na z^i
  -\ol\psi_A \l^{iA} \dx)\dx] \O^a + \nn\\
  &&-\frac{1}{4} \b_1 g_{i \bar \jmath}{\bf Z}^i_e {\bf \ol Z}^{\bar \jmath}_f
  \eta^{ef} \O + \nn\\
  &&+{\rm i} \b_2 g_{i\bar \jmath} \sx(\ol\l^{iA}\g^a\na\l^{\bar \jmath}_A +
  \ol \l^{\bar \jmath}_A \g^a \na\l^{iA} \dx) \O_a + \nn\\
  &&+{\rm i}\b_3\sx(\op{N}_{\L\S} {\bf \hat F}^{\L+}_{ab} + \ol \op{N}_{\L\S}
  {\bf \hat F}^{\L-}_{ab}\dx) \sx[\hat F^\S +\dx.\nn\\
  &&\sx.\quad-{\rm i}\sx(f_i^\S \ol \l^{iA} \g_c \psi^B
  \e_{AB}+\ol f_{\bar \imath}^\S \ol \l^{\bar \imath}_A \g_c \psi_B
  \e^{AB}\dx)  V^c\dx]  V^a  V^b+\nn\\
  &&-\frac{1}{24} \b_3 \sx(\ol\op{N}_{\L\S} {\bf \hat F}^{\L+}_{ef} {\bf
  \hat F}^{\S+~ef} - \op{N}_{\L\S} {\bf \hat F}^{\L-}_{ef}
  {\bf \hat F}^{\L-~ef} \dx) \O + \nn\\
  &&+d_1 M^{IJ} {\bf h}_{Id}\nn\\
  &&~~~~~\sx[ H_J -2\, {\rm i}\, \D_J{}^\a{}_\b\,\ol \z_\a \g^a \z^\b \O_a \dx.\nn\\
  &&~~~~~\sx.- \imez \sx(g_J{}^{A\a} \ol \psi_A \g_{ab} \z_\a - g_{JA\a} \ol
  \psi^A \g_{ab} \z^\a \dx)  V^a  V^b \dx]  V^d +\nn\\
  &&-\frac{1}{48}d_1 M^{IJ} {\bf h}_{Ie} {\bf h}_{Jf} \eta^{ef} \O +\nn\\
  && +b_1 {\bf P}_{aA\a} \sx(P^{A\a} -\ol \psi^A\z^\a -\e^{AB}
  \IC^{\a\b} \ol \psi_B\z_\b\dx) \O^a+\nn\\
  && -\frac{1}{8}b_1 {\bf P}_e{}^{A\a} {\bf P}^e{}_{A\a} \O +\nn\\
  &&+{\rm i}b_2 \sx( \ol \z^\a \g^a \na \z_\a + \ol \z_\a \g^a \na \z^\a \dx)
  \O_a +\nn\\
  &&+d_2 \sx(H_I -2\,{\rm i}\, \D_I{}^\a{}_\b\, \ol \z_\a \g^a \z^\b \O_a + \dx.\nn\\
  &&\sx.\qquad + \o_{IA}{}^B \, \ol\psi^A  \g_a \psi_B  V^a \dx)A^I{}_u
  P^u{}_{C\a}  P^{C\a}\\
  \label{eq:rhlagpauli}
  \op{L}_{\rm Pauli} &=& \b_5 \hat F^\L  \sx(\op{N}_{\L\S} L^\S \ol\psi^A  \psi^B
  \e_{AB} +\ol\op{N}_{\L\S}\ol L^\S\ol\psi_A  \psi_B \e^{AB} \dx)+\nn\\
  &&+{\rm i}\b_6 \hat F^\L  \sx(\ol\op{N}_{\L\S}f_i^\S \ol\l^{iA}\g_a\psi^B
  \e_{AB} + \op{N}_{\L\S} \ol f_{\bar \imath}^\S \ol\l^{\bar \imath}_A \g_a
  \psi_B \e^{AB} \dx) V^a + \nn\\
  &&+\b_7 \hat F^\L \sx(\op{N}-\ol\op{N}\dx)_{\L\S} \nn\\
  &&\quad \sx( \na_i f^\S_j \ol\l^{iA} \g_{ab} \l^{jB} \e_{AB} - \na_{\bar
  \imath} \ol f^\S_{\bar\jmath} \ol \l^{\bar\imath}_A \g_{ab} \l^{\bar\jmath}_B
  \e^{AB} \dx)  V^a  V^b +\nn\\
  &&+b_5 \hat F^\L \sx(\op{N}-\ol\op{N}\dx)_{\L\S} \sx( L^\S
  \ol \z_\a \g_{ab} \z_\b \IC^{\a\b} - \ol L^\S \ol \z^\a \g_{ab}
  \z^\b \IC_{\a\b} \dx)  V^a  V^b +\nn\\
  &&+b_3 \sx( P^{A\a} \ol \z_\a \g^{ab}  \psi_A +P_{A\a}\ol \z^\a \g^{ab}
   \psi^A \dx) \e_{abcd}  V^c  V^d +\nn\\
  &&+d_3 \sx( H_I -2\, {\rm i}\, \D_I{}^\a{}_\b \ol \z_\a \g^a \z^\b \O_a \dx)
  \sx( \op{U}^{IA\g} \ol \z_\g  \psi_A + \op{U}^I{}_{A\g} \ol
  \z^\g   \psi^A\dx) +\nn\\
  &&+d_4 \sx( H_I -2{\rm i} \D_I{}^\a{}_\b \ol \z_\a \g^a \z^\b
  \O_a \dx) \D^I{}_\g{}^\d\sx( \ol\z^\g \g_d \z_\d\dx)  V^d\\
  \label{eq:rhlagtop}
  \op{L}_{\rm top} &=& \a \, e^I_\L \sx[ \hat F^\L - L^\L \ol \psi^A  \psi^B \e_{AB}
  -\ol L^\L \ol\psi_A  \psi_B \e^{AB}- {\rm i} \sx( f^\L_i \l^{iA} \g_a \psi^B \e_{AB}+\dx.\dx.\nn\\
  &&\sx.\sx.\quad  +\ol
  f^\L_{\bar\imath} \l^{\bar\imath}_ A \g_a \psi_B \e^{AB} \dx)  V^a - m^{J\L}
  B_J\dx]  B_I\\
  \label{eq:rhlagtors}
  \op{L}_{\rm tors} &=& \sx(\b_4 g_{i\bar \jmath} \ol\l^{iA}\g_a
  \l^{\bar \jmath}_A + b_4 \ol \z^\a \g_a \z_\a \dx)T_b  V^b  V^a\\
  \label{eq:rhlaggauge}
  \op{L}_{\rm Shifts} &=& {\rm i} \d_1 \sx(S_{AB}\,\ol \psi^A  \g_{ab} \psi^B +
  \ol S^{AB} \,\ol \psi_A  \g_{ab} \psi_B \dx)  V^a  V^b + \nn\\
  && +{\rm i}\, \d_2\, g_{i\bar \jmath} \sx( W^{iAB}\, \ol\l^{\bar\jmath}_A
  \g^a \psi_B +W^{\bar \jmath AB}\, \ol \l^{iA} \g^a \psi^B \dx) \O_a + \nn\\
  && +{\rm i}\,\d_3 \sx(N_A^\a\, \ol\z_\a \g^a \psi^A +N^A_\a\, \ol\z^\a \g^a
  \psi_A \dx) \O_a +\nn\\
  && +\sx(\d_4 \na_u N_A^\a P^u{}_{B\b} \e^{AB} \IC^{\a\b}\ol\z_\a\z_\g
  +\d_5 \na_i N_A^\a \ol \z_\a\l^{iA} +\dx.\nn\\
  &&\quad \sx. +\d_6 g_{i\bar \jmath} \na_k W^{\bar \jmath}_{AB}\ol\l^{iA}
  \l^{kB}+\cc \dx) \O \\
  \op{L}_{\rm Potential} &=& - \frac{1}{6} \op{V}(q, z, \ol z) \O \label{eq:rhlagpot}
\end{eqnarray}

\noi where we have defined

\begin{equation}
\O = \e_{abcd} V^a V^b V^c V^d,\quad \O_a = \e_{abcd} V^b V^c V^d
\end{equation}

\noi and the scalar potential $\op{V}(q, z, \ol z)$ is given by
equation (\ref{poten}). Furthermore we have introduced auxiliary
0--forms ${\bf \hat F}^{\pm\L}_{ab}$, ${\bf Z}^i_a$, ${\bf
h}_{Ia}$, ${\bf P}_a{}^{A\a}$ whose variational equations identify
them with ${\tilde{ \hat F}}^{\pm\L}_{ab}$, ${\tilde Z}^i_a$,
${\tilde h_{Ia}}$, ${\tilde P}_a{}^{A\a}$ defined by the solution
of the Bianchi identities. These auxiliary fields have to be
introduced in the kinetic terms of the Lagrangian in order to
avoid the Hodge duality operator which would destroy the
independence of the variational equation from the particular
bosonic hypersurface of integration.\\
The variational equations, together with the principles of
rheonomy fix the undetermined coefficients in the Lagrangian to
the following values:

\begin{eqnarray}
&& \b_1 = \frac{2}{3};\quad \b_2 = - \frac{1}{3};\quad \b_3 = 4
{\rm i};\quad
\b_4= -1;\quad \b_5 = 4;\quad \b_6 = - 4;\nn\\
&& \b_7 =\frac{1}{2};\quad b_1 = -\frac{4}{3};\quad b_2 =
\frac{2}{3};\quad b_3 = 2;\quad b_4 = - 2;\quad b_5=1;\nn\\
&& \d_1 =4;\quad \d_2 =\frac{2}{3};\quad \d_3 = -\frac{4}{3};\quad
\d_4=- \frac{1}{12};\quad \d_5=-\frac{1}{3};\quad \d_6=\frac{1}{18};\nn\\
&& d_1=-8;\quad d_2=8;\quad d_3=-8;\quad d_4=-8{\rm i}; \quad
\a=8.
\end{eqnarray}\label{coeff}

\noi In order to obtain the space--time Lagrangian the last step
to perform is the restriction of the 4--form Lagrangian from
superspace to space--time. Namely we restrict all the terms to the
$\th = 0\,,\,{\rm d} \th = 0$ hypersurface ${\cal M}^4$. In
practice one first goes to the second order formalism by
identifying the auxiliary 0--form fields as explained before. Then
one expands all the forms along the ${\rm d}x^{\mu}$ differentials
and restricts the superfields to their lowest ($\theta = 0$)
component. Finally the coefficient of:

\begin{equation}
dx^{\mu}\wedge dx^{\nu}\wedge dx^{\rho}\wedge
dx^{\sigma}\,=\,{\epsilon^{\mu\nu\rho\sigma}\over \sqrt g}\left(
\sqrt{g}\, {\rm d}^4x \right)
\end{equation}

\noi gives the Lagrangian density written in section
\ref{Lagrangian}. The overall normalization of the space--time
action has been chosen such as to be the standard one for the
Einstein term.

\end{document}